\documentclass[a4paper,onecolumn,11pt]{quantumarticle}
\pdfoutput=1
\usepackage[utf8]{inputenc}
\usepackage[english]{babel}
\usepackage[T1]{fontenc}
\usepackage{amsmath}
\usepackage{amssymb}
\usepackage{hyperref}

\usepackage{tikz}
\usepackage{lipsum}

\newtheorem{lemma}{Lemma}
\newtheorem{proof}{Proof}

\usepackage[numbers,sort&compress]{natbib}
\begin{document}

\title{ Finite-key analysis for  quantum key distribution with discrete phase randomization}

\author{ Rui-Qiang Wang}

\author{Zhen-Qiang Yin}
\email{yinzq@ustc.edu.cn}
\affiliation{CAS Key Laboratory of Quantum Information, University of Science and Technology of China, Hefei 230026, P. R. China}
\affiliation{Synergetic Innovation Center of Quantum Information $\&$ Quantum Physics, University of Science and Technology of China, Hefei, Anhui 230026, P. R. China}
\affiliation{State Key Laboratory of Cryptology, P. O. Box 5159, Beijing 100878, P. R. China}

\author{Rong Wang}
\affiliation{Department of Physics, University of Hong Kong, Pokfulam, Hong Kong }
\author{Shuang Wang}
\author{Wei Chen}
\author{Guang-Can Guo}
\author{Zheng-Fu Han}
\affiliation{CAS Key Laboratory of Quantum Information, University of Science and Technology of China, Hefei 230026, P. R. China}
\affiliation{Synergetic Innovation Center of Quantum Information $\&$ Quantum Physics, University of Science and Technology of China, Hefei, Anhui 230026, P. R. China}
\affiliation{State Key Laboratory of Cryptology, P. O. Box 5159, Beijing 100878, P. R. China}

\maketitle

\begin{abstract}
  Quantum key distribution(QKD) allows two remote parties to share information-theoretic secret keys. Many QKD protocols assume the phase of encoding state can be continuous randomized from $0$ to $2\pi$, which, however, may be questionable in experiment. This is particularly the case in the recently proposed twin-field(TF) QKD, which has received a lot of attention, since it can increase key rate significantly and even beat some theoretical rate-loss limits. As an intuitive solution, one may introduce discrete phase-randomization instead of continuous one.  
However, a security proof for a QKD protocol with discrete phase-randomization in finite-key region is still missing. Here we develop a technique based on conjugate measurement and quantum state distinguishment to analyze the security in this case. Our result shows that TF-QKD with reasonable number of discrete random phases, e.g. 8 phases from $\{0,\pi/4,\pi/2,...,7\pi/4\}$,  can achieve satisfactory performance.
More importantly, as a the first proof for TF-QKD with discrete phase-randomization in finite-key region, our method is also applicable in other QKD protocols.  
\end{abstract}

\section{introduction}
Quantum key distribution(QKD) \cite{ekert1991quantum,bennett1984proceedings}, the most successful and mature application in quantum  information science, allows two legtimate parties Alice and Bob to share information-theoretic secret keys. In theory, its security has been proved \cite{10.1145/382780.382781,lo1999unconditional,shor2000simple}, while experiments towards higher key rate\cite{dixon2015high} and longer acheivable distance\cite{yin2016measurement,liao2017satellite,boaron2018secure} have been demonstrated. Even, some large scale QKD networks are emerging\cite{poppe2008outline,sasaki2011field,wang2014field,chen2021integrated}. However, owing to the  inherent photon-loss in the channel, it meets a vital bottleneck which limits the communication distance and key generation rate. Specifically, some fundamental rate-loss limits \cite{takeoka2014fundamental,pirandola2017fundamental} impose restriction on any point-to-point QKD without repeaters. For instance, the key rate $R$ is bounded by the channel transmission probability $\eta$ with the linear bound $R = -\log_2(1-\eta)$\cite{pirandola2017fundamental}. Delightfully, M.Lucamarini et al. made a breakthrough by proposing twin-field(TF) QKD in 2018. The essential idea of TF-QKD is in code mode extracting key bit from single photon click event of the measurement station located in the middle of channel, which happens with a probability proportional to $\sqrt{\eta}$, thus surpassing linear bound becomes possible; a so-called phase error rate may be estimated in decoy mode \cite{hwang2003quantum,lo2005decoy,wang2005beating} to monitor security. Driven by this, several TF-type QKD protocols \cite{tamaki2018information,ma2018phase,curty2019simple,wang2018twin,lin2018simple,yin2019measurement,wang2020optimized} were proposed later to complete security proofs and improve its performance. Based on these protocols, experimentalists also made great efforts to realize TF-QKD \cite{minder2019experimental,pittaluga2021600,chen2021twin,wang2019beating,zhong2019proof,fang2020implementation,liu2019experimental,chen2020sending}.

Since TF-QKD inherits measurement-device-independent(MDI)-QKD's \cite{lo2012measurement}merit that is immune to all side-channel attacks to measurement devices, one doesn't need to take the detection loopholes into account within TF-type QKD system. In spite of this, the security issues of the state preparation in TF-QKD must be carefully considered. In practice, the laser source of TF-QKD is usually a continuous source emitting coherent states with a fixed phase. Meanwhile continuous phase-randomization from $0$ to $\pi$ is required in the TF-QKD. More specifically, this continuous phase-randomization is assumed in both code and test modes in Refs.\cite{ma2018phase, wang2018twin,wang2020optimized}, or at least test mode in Refs.\cite{cui2019twin,lin2018simple,curty2019simple}. To fulfill this requirement, Alice and Bob must randomize the global phase continuously and uniformly. Unluckily, two ways to achieve phase-radomization introduce different problems in experiment. Passive randomization will lead to phase correlations between adjacent pulses \cite{xu2012ultrafast,abellan2014ultra} while active randomization can only randomise the phase over discrete set of values. 

To bridge this gap between theory and experiment, two works which analyzed the security of fully discrete-phase-randomization TF-QKD protocol have been proposed \cite{PhysRevApplied.14.064070,curras2020twin} in these days. However, a security proof in finite-key region is still missing.
 Hence, one natural question is that if TF-QKD with fully discrete-phase-randomization can work well non-asymptotically.  This work gives the affirmatory answer. 

In this paper, we analyze the security of TF-QKD protocol with fully discrete randomization in finite-key region.
Interestingly,   our analysis leads to comparable performance with the continuous one.  Since taking the discrete phase into account, our results make the TF-QKD more practical and  can be applied to future TF-QKD experiment. More importantly, some techniques proposed here, e.g. Lemma 1 (introduced later) can be utilized to analyze the security of other QKD protocols with discrete phase randomization.

 This work is organized as follows. In section 2, we give a description of  TF-QKD protocol with fully discrete phase randomization, and the sketch of  security proof is given in section 3. Note the proof is detailed in Appendix.
 In section 4, by the numerical simulation,
 we show this protocol  can still beat the linear bound \cite{pirandola2017fundamental} and has satisfactory performance.  Finally, a conclusion is given in section 5.

\section{Protocol description}
Indeed the protocol analyzed here has been depicted in Ref.\cite{curras2020twin}. For ease of understanding, we illustrate the protocol as follows. 

Step 1: Alice(Bob) chooses a label from $\{"\mu","0","\nu" \}$ with probabilities $ P_{\mu},P_O,P_{\nu}$, according to the label she(he) chooses, she(he) takes one of the following actions:

\quad $"\mu"$: She(he) randomly picks an integer $l_{A_c}$ ($l_{B_c}$)  from $\{ 0,1,\cdots, M-1\}$ with equal probability $\frac{1}{M}$ where M is an even integer. It means that the  phase 2$\pi$ is divided into M parts. Then, she(he) randomly chooses a  key bit $k_a$$(k_b)$ where $k_a (k_b) \in \left\{ 0,1\right\} $. Finally she(he) sends a pulse with coherent state $|e^{i(\frac{l_{Ac}}{M}2\pi + \pi k_a)}\sqrt{\mu}\rangle $($|e^{i(\frac{l_{B_c}}{M}2\pi + \pi k_b)}\sqrt{\mu}\rangle $ ). 

\quad $"0"$:  She(he) sends the vacumm state.

\quad $"\nu"$: She(he) randomly picks an integer $l_{A_c}$ and $l_{B_c}$  from $\{ 0,1,\cdots, M-1\}$ with equal probability $\frac{1}{M}$ where M is an even integer. It means that the  phase 2$\pi$ is divided into M parts. Then, she(he) sends a pulse with coherent state $|e^{i\frac{l_{Ac}}{M}2\pi}\sqrt{\mu}\rangle $($|e^{i\frac{l_{Bc}}{M}2\pi }\sqrt{\mu}\rangle $ ).

The first case is called code mode, while the other cases are decoy mode.

Step 2: Alice and Bob repeat Step 1 in total of $N_{tot}$ times.

Step 3: After receiving $N_{tot}$ pairs of pulses from Alice and Bob, interfering each pair at a beamsplitter and measuring the two outputs with his single photon detectors(SPDs), an honest Eve announces whether or not each measurement is successful. Here 'successful' means only one SPD (left SPD or right SPD) clicks in the corresponding measurement, and if so, Eve reports the specific SPD clicked.

Step 4: For those rounds Eve announcing successful click, Alice and Bob announce the intensities they chose as well as the values of $l_{A_c}$ and $l_{B_c}$. Then Alice and Bob only retain those successful rounds in which the intensities of the coherent state they sent are same while in-phase ($l_{A_c}=l_{B_c}$) or anti-phase  ($|l_{A_c}-l_{B_c}|=M/2$) condition is also met.  Let $n_{2\beta}^{+}$($n_{2\beta}^{-}$) be the number of the retained rounds when both Alice and Bob chose the same intensity $\beta$ of coherent state and in-phase(anti-phase) is also met. Note that we assume $l_{A_c}=l_{B_c}=0$ always holds in case of $\beta=0$. Alice and Bob generate their sifted keys from $n_{2\mu}=n^+_{2\mu}+n^-_{2\mu}$ retained rounds in code mode, thus the length of sifted key bits $n_{bit}=n_{2\mu}$. Note that if it's a in-phase(anti-phase) round with right(left) SPD clicking, Bob may flip his corresponding sifted key bit.

Step 5:  With all quantities $n_{2\beta}=n^+_{2\beta}+n^-_{2\beta}$, Alice and Bob  use linear programming to get an upper bound on the number of phase errors(defined later) $n^U_{ph}$ with a failure probability no more than $\varepsilon$, then they can calculate the upper bound $ e^{U}_{ph} = n^{U}_{ph}/n_{bit} $.    

Step 6: Step 6 consists of error correction and privacy amplification.

\quad Step 6a: Alice sends $H_{EC}$ bits of syndrome information of her sifted key bits to Bob through an authenticated public channel. Then Bob uses it to correct errors in his sifted keys. Alice and Bob calculate a hash of their error-corrected keys with a random universal hash function, and check whether they are equal. If equal, they continue to the next step, otherwise, they abort the protocol

\quad Step 6b: Alice and Bob apply the privacy amplification to obtain their final secret keys. If the length of their secret key satisfies  $ l = n_{bit}(1-h(e^U_{ph} )) - H_{EC} - \log_2\frac{2}{\epsilon_{cor}} - \log_2 \frac{1}{4\epsilon_{PA}^2}$ where $h(\cdot)$ denotes the binary Shannon entropy, this protocol must be $\epsilon_{cor}$-correct and $\epsilon_{sec}$-secret with $ \epsilon_{sec}=  \sqrt {\varepsilon} + \epsilon_{PA}$. Here $\epsilon_{cor}$($\epsilon_{sec}$) represents the protocol is correct(secret) with a failure probability no more than $\epsilon_{cor}(\epsilon_{sec})$.
 Hence, the total security parameter is $\epsilon_{tol}$-secure where $\epsilon_{tol} = \epsilon_{cor} +\epsilon_{sec}$. It is elaborated thoroughly in the widely-used universally composable security framework \cite{ben2005universal,muller2009composability}

\section{security proof}
In this section, we present the security proof of this protocol. The main task of security proof is to bound the information Eve holds. To accomplish this task, one can calculate a so-called phase error rate. Firstly, we construct an equivalent virtual protocol, in which Alice and Bob prepare some entangled states between local states and traveling states, but traveling states must have the same density matrices as actual protocol in the channel. The sifted key bits can be seen as the outputs of measurement with $Z$-basis on local states made by Alice and Bob, then the so-called phase error rate is defined as the error rate for the outputs of measurement with $X$-basis made by them. According to the  complementarity argument\cite{koashi2009simple}, the phase-error rate can be used to bound Eve's information on the sifted keys. In the following, we give the virtual protocol and show how to bound the phase error rate.
\subsection{ Equivalent virtual protocol}
\label{equivalent protocol}
In our virtual protocol, Alice generates secret keys from code mode in which she prepares state
\begin{eqnarray}
|\psi\rangle_{\mu,A_c A a} = \sum_{l = 0}^{M -1} \frac{1}{\sqrt{M}} |l\rangle_{A_c} ( \frac{1}{\sqrt{2}}(|0\rangle_A |e^{i\frac{2\pi}{M}l} \sqrt{\mu}\rangle_a + |1\rangle_A|-e^{i\frac{2\pi}{M}l}\sqrt{\mu} \rangle_a) ),
\end{eqnarray}
where $A_c$ and A are  the local quantum systems in Alice's side, $a$ is the traveling quantum state Alice sent to Eve, Similarily, Bob prepares $ |\psi\rangle_{\mu,B_c B b} $ defined analogously to $|\psi\rangle_{\mu,A_c A a}$. Obviously,  Alice(Bob) measures $A(B)$ with $Z$-basis to obtain sifted key, i.e. $|0\rangle_A$ for bit $0$ and $|1\rangle_A$ for bit $1$. In order to get the phase-error rate, they measure $A,B$ in $X$-basis $\{   |+\rangle ,|-\rangle \}$ after Eve's attack.  As for the test mode, we assume Alice prepares the following states
\begin{align}
   |\psi\rangle _{0,A_c A a} &=  |0\rangle_{A_c}|0\rangle_A |0\rangle_a, \\ \notag
   |\psi\rangle_{\nu,A_c A a} &= \sum_{l=0}^{M-1} \frac{1}{\sqrt{M}} |l\rangle_{A_c}   |0\rangle_A |e^{i\frac{2\pi}{M}l} \sqrt{\nu}\rangle_a .  \\ \notag 
\end{align}
Here the local states of $A_c$ is encoded in photon-number states and  Alice can measure $A_c$'s photon-number to learn the phase of sent  states.
 
Finally, we can describe the process of state preparation above with a single state, namely
\begin{eqnarray}
|\psi\rangle_{A_s A_c Aa} = \sqrt{p_{\mu}}|0\rangle_{A_s} |\psi\rangle_{\mu,A_c A a} + \sqrt{p_O} |1\rangle_{A_s} |\psi\rangle_{0,A_c A a} + \sqrt{p_{\nu}}|2\rangle_{A_s}|\psi\rangle_{\nu,A_c A a} 
\end{eqnarray}
where Alice's additional local ancilla  $A_s$ is in the photon number states. Similarly, Bob can prepare $|\psi\rangle_{B_s B_cBb}$ defined analogously to $|\psi\rangle_{A_s A_c Aa}$. Though Alice (Bob) may measure $A_s$($B_s$),$A_c(B_c)$ and $A$($B$) after or before Eve announcing her measurement results, Alice(Bob) must announce the measurement results after Eve's announcement then post-selects the successful rounds.
The following is a detailed illustration of our equivalent virtual protocol. 

Step 1: 

\quad Alice and Bob prepare a gigantic quantum state $|\Phi\rangle = |\phi\rangle^{\otimes N_{tot}}=( |\psi\rangle_{A_sA_cAa} \otimes |\psi\rangle_{B_sB_cBb} )^{\otimes N_{tot}} $ and send all subsystems $a$ and $b$ to Eve through an insecure quantum channel.

Step 2:

\quad After performing arbitrary quantum operation on all subsystems $a$ and $b$ from Alice and Bob, Eve  announces whether it has a successful click (only one of her SPDs clicks) or not for each round. For a successful round, Eve continues to announce whether the left SPD clicks or the right SPD clicks. We use $\mathcal{M}$($ \overline{\mathcal{M}}$) to denote the set of successful(unsuccessful) rounds.

Step 3:

\quad For those rounds Eve announcing success, Alice and Bob jointly measure the subsystem $A_c$($B_c$)  and  $A_s(B_s)$ in the photon-number basis to learn whether the intensities of coherent state they send are same or not and whether it is in-phase or anti-phase. Then they only retain those rounds where in-phase or anti-phase is met, and they chose the same intensities. Let $\mathcal{M}_s$ denote the set of those retained rounds, while $\mathcal{M}_f $ denotes those rounds which are in $\mathcal{M}$ but not in $\mathcal{M}_s$.

Step 4:

 For these rounds in $\mathcal{M}_s$, Alice(Bob) measures the subsystem $A_cA_s $($B_cB_s $) in Fock basis to learn the phase and intensity of the coherent states they sent. If the result of   $A_s$($B_s$) is in state $|0\rangle_{A_s}$($ |0\rangle_{B_s}$), she(he) measures subsystems $A$($B$) in the Z basis to decide her(his) sifted key respectively, otherwise, she(he) measures subsystem A(B) in the Z basis but doesn't incorporate these measurement outcomes in her(his) sifted key.

Step 5 to Step 6:

\quad Let $n_{2\beta}$ be the number of rounds in $\mathcal{M}_s$ satisfying that  both Alice and Bob chose the intensity $\beta$.  With  parameters $n_{2\beta}$, perform the same operations as Step 5 to Step 6 respectively in the actual protocol given in section II.

\subsection{ Estimation of phase-error rate } 
The essential of security proof is to estimate the upper-bound of phase error rate $e_{ph}$ of the sifted keys, i.e. how many same or different outcomes Alice and Bob have if they measure $A$ and $B$ with $X$-basis hypothetically in the rounds where sifted keys are generated. Specifically, in our protocol, we define the number of the same outcomes they have as $n_{ph}$, i.e. the number of phase error events.  Provided that $e_{ph}=n_{ph}/n_{2\mu}$ is bounded,  one can generate final secret key with appropriate  $\epsilon_{tol}$ value as given in Step6.b of the actual protocol.

A detailed proof for how to estimate $n_{ph}$ is present in Appendix. Here, a sketch of this proof is given.

Though analyzing the equivalent protocol, it's proven that if Alice and Bob both chose intensity $\beta$, and in-phase or anti-phase is also met, they actually prepare a mixture $\tau_{2\beta}$,  which consists of  component $\tau_{j|2\beta},j=0,1,...,M-1$.  Moreover, each phase error event is a click by some particular components of that mixture $\tau_{2\beta}$, i.e. $\tau_{j|2\beta},j=0,2,...,M-2$. These results imply that 
\begin{eqnarray}
	n_{2\mu} =& \sum_{j=0}^{M} n_{j|2\mu}, \notag \\
	n_{2\nu} =& \sum_{j=0}^{M} n_{j|2\nu}, \notag \\
	n_{ph} =& \sum_{j=0,j\in \mathcal{N}_0}^{M-2} n_{j|2\mu},
	\label{eq:1}
\end{eqnarray} 
where $n_{j|2\beta}$ denotes the number of rounds in $\mathcal{M}_s$, in which Alice and Bob both chose intensity $\beta$, but $\tau_{2\beta}$ is actually $\tau_{j|2\beta}$. Meanwhile, $\mathcal{N}_0$ is the set of even number. Now, the hypothetical value $n_{ph}$ is related on some experimentally observed values. Howevere, just with these equations, it's difficult to bound $n_{ph}$ tightly, since $n_{j|2\beta}$ cannot be known directly. 

On the other hand,  both $\tau_{j|2\mu}$ and $\tau_{j|2\nu}$ are very close to Fock-state $|j\rangle\langle j|$. Accordingly, it's intuitive to consider there are constraints on the gap between $n_{j|2\mu}$ and $n_{j|2\nu}$. Then we developed Lemma1 (see appendix for details) to bound the gap between yields of two distinct quantum states in non-asymptotic situation. Applying this lemma, we obtain a series of constraints on $n_{j|2\mu}$ and $n_{j|2\nu}$. Finally, combined with Eq \eqref{eq:1}, a linear programming (given in the end of the Appendix) is introduced to find the upper bound of  phase error rate $e^U_{ph}=n_{ph}/n_{2\mu}$.

 \section{Numerical simulation} 
In this section, we simulate the final secret key rate with the parameters listed in TABLE \ref{table:1}.
\begin{table}[htbp]
	\centering
	\begin{tabular}{ccccccc}
		\hline 
		$e_m$ & $p_d$ &  $\xi$(dB/km) & $\eta_d$ &$f$&  $\epsilon_{tol}$ &   \\
		\hline
		0.03 & $1\times10^{-8}$ & 0.2 & 0.3& 1.1 &$4.6566\times 10^{-10}$  \\
		\hline 	
	\end{tabular}
	\caption{\label{table:1}List of parameters uesd in the numerical simulations. Here, $e_m$ is loss-independent misalignment error rate due to optical imperfect interference, $p_d$ is dark counting probability for each SPD, $\xi$ is fiber loss constant, $\eta_d$ denotes detection efficiency of each SPD, $f$ is error-correction inefficiency and $\epsilon_{tol}$ denotes the total security coefficient. }
	\label{table}
\end{table} 
It's reasonable to simulate the experimentally observed values $n_{2\mu}$, $n_{2\nu}$ and $n_0$ with their mean values respectively. Let $Q_{corr|2\beta}$ be the probability of  only one click from left (right) SPD when both Alice and Bob prepare coherent states with intensity $\beta$ and phase difference of $0$ ($\pi$), and $Q_{err|2\beta}$ be the probability of  only one click from left (right) SPD when both Alice and Bob prepare coherent states with intensity $\beta$ and phase difference of $\pi$ ($0$). Then we have

\begin{eqnarray}
    Q_{corr|2\beta}=(1-(1-p_d)e^{-2\eta(1-e_m)\beta})e^{-2\eta e_m\beta}(1-p_d), \notag \\
	Q_{err|2\beta}=(1-(1-p_d)e^{-2\eta e_m\beta})e^{-2\eta (1-e_m)\beta}(1-p_d),
\end{eqnarray}
where $\eta = 10^{\frac{-0.2L}{20}}$ and $L$ is the channel distance between Alice and Bob.
Accordingly, in the simulation, we assume $n_{2\beta}=N_{tot}P^2_\beta 2 (Q_{corr|2\beta}+Q_{err|2\beta})/M$ for $\beta=\mu,\nu$. Note that $n_0=N_{tot}P^2_O (Q_{corr|0}+Q_{err|0})$, $n_{bit}=n_{2\mu}$ and $e_{bit}=Q_{err|2\mu}/(Q_{corr|2\mu}+Q_{err|2\mu})$.  With these values, setting $M=8$ and the failure probability of estimating phase error  $\varepsilon= (8M+12)\varepsilon_a =4\times 10^{-20}$, one can obtain the upper-bound of phase error rate $e^U_{ph}$ by the linear programming given by \eqref{linearprogramming} in Appendix. Moreover, the amount of $H_{EC}$ is $H_{EC} = N_{bit}f h(e_{bit})$, $\epsilon_{cor} = 1 \times 10^{-10}
$ and $\epsilon_{PA}= 1.6566\times 10^{-10}$, which leads to a secret key of length  $ l = n_{bit}(1-h(e^U_{ph} )) - H_{EC} - \log_2\frac{2}{\epsilon_{cor}} - \log_2 \frac{1}{4\epsilon_{PA}^2}$ with $\epsilon_{sec}=\epsilon_{PA}+ \sqrt{\varepsilon}$ and the total security parameter $\epsilon_{tol}=\epsilon_{cor}+\epsilon_{sec}= 4.6566 \times 10^{-10}$.

 Finally, we numerically optimize the intensities and corresponding probabilities to maximize $l$ in the cases of the total number of pulses is $N_{tot}=1\times10^{14},1\times10^{13},1\times10^{12}$.  The simulate results ($l/N_{tot}$ v.s. L)  are illustrated below.
 
  \begin{figure}[!htbp]
  \includegraphics{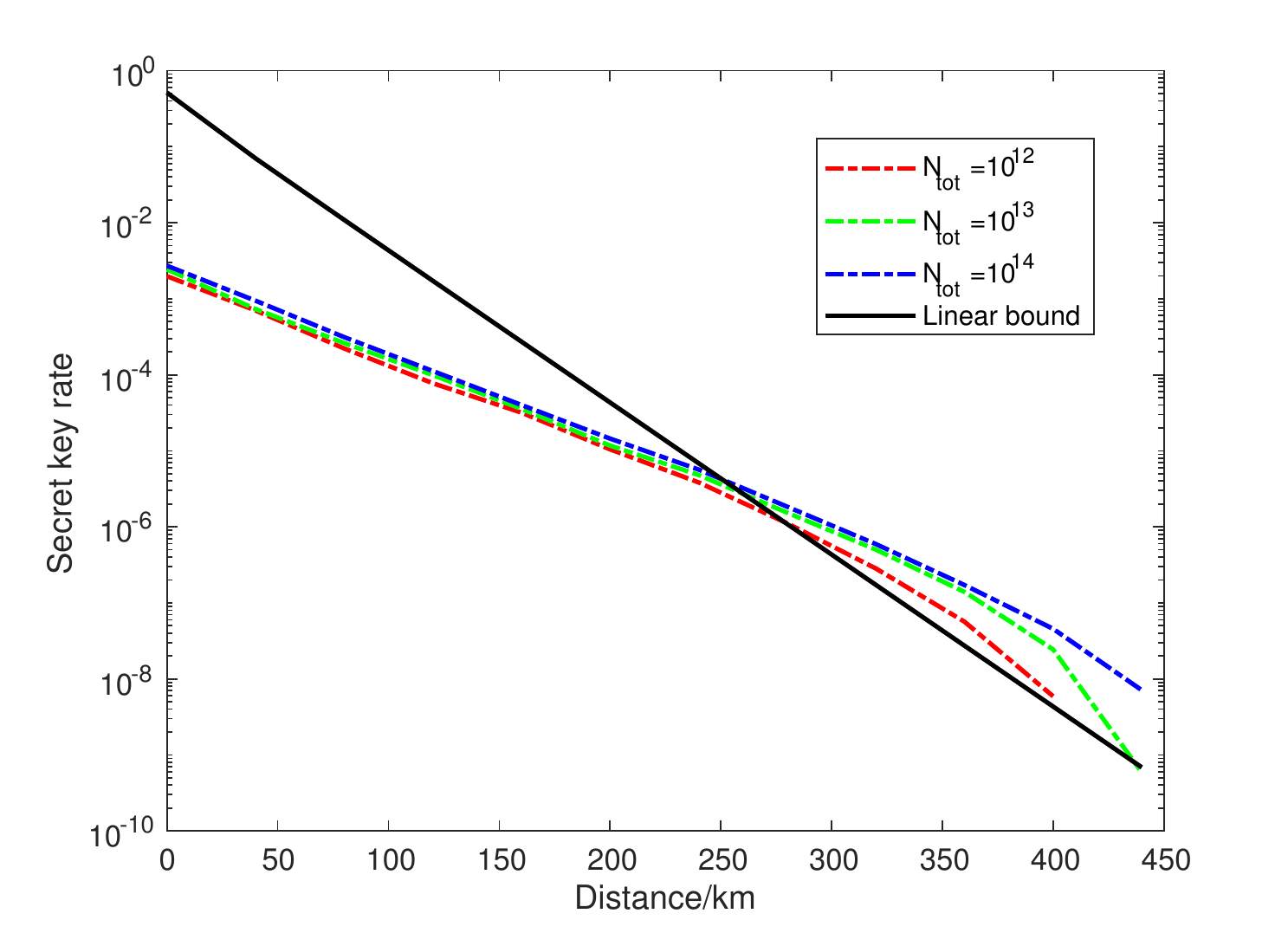}%
  \caption{\label{fig:1}Secret key rate ($l/N_{tot}$) of fully discrete TF-QKD\cite{ curras2020twin}. In this figure, the key rate corresponding to the total number of pulses $N_{tot}$ is $1\times10^{12},1\times10^{13},1\times10^{14}$ are plotted. Note that we set $M=8$ in the simulation. Obviously, the linear bound\cite{pirandola2017fundamental}} is broken at about 250 km.
  \end{figure}

As  Fig.\ref{fig:1}  shows, we get considerable secret key rates  when the total number of pulses is  $10^{12}$, $10^{13}$ or $10^{14}$. Furthermore, it's clear that the linear bound \cite{pirandola2017fundamental} can also be overcome. Hence, we make a meaningful breakthrough that we not only make the TF-QKD more practical but also keep its performance.  

\section{Conclusion}
In real setups of TF-QKD, the continuous randomization is usually realized by actively adding a random signal to a phase modulator. On the other hand, random numbers are generated discretely in most schemes. Therefore, TF-QKD with discrete phase randomization is more practical. It is necessary to analyze the security of TF-QKD with discrete phase randomization. Based on conjugate measurement, the security proof of a QKD protocol is to estimate the phase error rate. Then in case of discrete phase randomization, a critical step is how to bound the gap between yields of two distinct but very close quantum states in non-asymptotic situation. To achieve this goal, Lemma 1 is developed to find the upper bound of this gap. With the help of Lemma 1, a linear programming is proposed to calculate phase error rate, and the key length is then straightforward. Through numerical simulations, it's confirmed that TF-QKD with discrete phase randomization has comparable performance with the continuous one. 

Besides, it's worth noting that Lemma 1 is quite useful in a variety of scenarios, not just in the security proof of TF-QKD. For instance, if one considers the BB84 with discrete phase randomization \cite{cao2015discrete}, the Lemma 1 can be utilized to bound the yield of single photon state, then it's not difficult to give a relevant security proof.
To summarize, we give the first security proof for TF-QKD with finite discrete phase randomization in non-asymptotic scenarios.  Although the proof is tailored for TF-QKD, the framework of this proof, i.e. Lemma 1, can be adapted in other protocols.

\section{Acknowledgments}
This work has been supported by the National Key Research and Development Program of China (Grant No.
2020YFA0309802), the National Natural Science Foundation of China (Grant Nos. 62171424, 61961136004, 61775207, 61627820) and Anhui Initiative in Quantum Information Technologies.

\onecolumn\newpage
\appendix

\section{ Appendix}
\subsection{ Formula for the number of phase error events}
In this section, we show how to obtain the relation between the hypothetical phase error events and some experimental observations. 

The main result obtained here is that each key bit is a successful click from a mixed  state of  $\tau_{j|2\mu},j=0,1,...,M-1$ prepared by Alice and Bob. More importantly,  the number of phase error events among these key bits correspond to   $\tau_{j|2\mu},j=0,2,...,M-2$. Therefore, if we denote the length of raw key by $n_{bit}=n_{2\mu}$ and the number of successful clicks of  $\tau_{j|2\mu}$ by $n_{j|2\mu}$, $n_{bit}=n_{2\mu}=\sum^{M-1}_{j=0}n_{j|2\mu}$ and the number of phase error events $n_{ph}=\sum_{j=0,j\in \mathcal{N}_0}^{M-2}n_{j|2\mu}$ must hold, where $\mathcal{N}_0$ is the set of even integers. Next, a proof is present to show how to get this result.

 Following the symbols in \cite{curras2021tight}, let us consider the evolution of the gigantic quantum state $|\Phi\rangle = |\phi\rangle^{\otimes N_{tot}}=( |\psi\rangle_{A_sA_cAa} \otimes |\psi\rangle_{B_sB_cBb} )^{\otimes N_{tot}} $ sent to Eve. After Step 2, where Eve performs her measurement on the subsystem $ab$, the initial quantum state is transformed to $\hat{M}_{eve} |\Phi\rangle $ where $\hat{M}_{eve}$ denotes the measurement operator of Eve. After measurement, Eve announces the measurement outcome is susccessful or not for each rounds, Hence, we reorder the quantum state as $|\Phi\rangle  = |\phi\rangle^{\otimes M} |\phi \rangle ^{\otimes \overline{M}}$  where $M(\overline{M}) $ denotes the  successful (unsuccessful) rounds. Then, in Step 3 of the virtual protocol, using measurement operators $\{ \hat{O}_s = (|00\rangle_{A_sB_s}  \langle 00| + |11\rangle_{A_sB_s}  \langle 11| + |22\rangle_{A_sB_s}  \langle 22|  )\otimes \sum_{l=0}^{M-1} (|l,l\rangle_{A_cB_c} \langle l,l |+|l,(l+M/2)mod M\rangle_{A_cB_c} \langle l,(l+M/2)mod M |) , \hat{O}_d = \hat{I} - \hat{O}_s \} $,  Alice and Bob measure the subsystem $A_sA_c$ and $B_s B_c$ for those rounds which are announced successful in Step 2 and retain the trials in which $A_sA_c$ and $B_s B_c$ are collapsed into $\hat{O}_s$ as the final successful rounds. Hence, we reorder $|\Phi\rangle = |\phi\rangle^{\otimes M_s}|\phi\rangle^{\otimes M_f }|\phi \rangle^{\otimes \overline{M} } $ where $M_s$($M_f$) denotes the successful(unsuccessful) rounds finally. Before they measure the subsystem $AB$ to generate their sifted key in Step 3, the unnormalised quantum state is given by

\begin{eqnarray}
\hat{O}_s^{M_s}\hat{O}_d^{M_f} \hat{I}^{\otimes \overline{M}} \hat{M}_{eve} |\Phi\rangle  = \hat{M}_{eve}\hat{O}_s^{\otimes M_s}\hat{O}_d^{\otimes M_f} \hat{I}^{\otimes \overline{M}}|\Phi\rangle = \hat{M}_{eve} (\hat{O}_s |\phi\rangle)^{\otimes M_s} (\hat{O}_f |\phi\rangle)^{\otimes M_f}  (|\phi\rangle)^{\otimes \overline{M}} 
\end{eqnarray}

Nextly, in Step 4, Alice and Bob measure the subsystem $A_s,B_s$ $A_c,B_c$ and $A,B$ for all rounds in $\mathcal{M}_s$, one by one. We use $\alpha \in \{1,\cdots,M_s\}$ to denote the different rounds in $\mathcal{M}_s$ and $\xi_{\alpha}$ to denote the measurement outcome of the $\alpha$-th subsystem.  What's more, $\hat{M}_{\alpha}$ is uesd to denote the associated operator. Hence, the unnormalised state before the measurement of the $\alpha$-th  rounds in $\mathcal{M}_s$ is
\begin{eqnarray}
|\Phi_{\alpha}\rangle = \hat{M}_{eve} (\otimes_{l=1}^{\alpha -1} \hat{M}_l |\phi\rangle)(\hat{O}_s |\phi\rangle)(\hat{O}_s |\phi\rangle)^{\otimes M_s-\alpha } (\hat{O}_f |\phi\rangle)^{\otimes M_f}  (|\phi\rangle)^{\otimes \overline{M}}.  
\end{eqnarray}
Because we are only interested in the reduced state of the $\alpha$-th round in $\mathcal{M}_s$, we trace out the other rounds which we denote by $\overline{\alpha}$ and get
\begin{eqnarray}
\hat{\sigma}_{\alpha} = Tr_{\overline{\alpha}} \left[ |\Phi_{\alpha}\rangle \langle |\Phi_{\alpha}|\right] = \sum_{\overrightarrow{\overline{\alpha}}} \langle \overrightarrow{\overline{\alpha}} |\Phi_{\alpha}\rangle \langle \Phi_{\alpha}| \overrightarrow{\overline{\alpha}} \rangle  = \sum_{\overrightarrow{\overline{\alpha}}}     \hat{M}_{\overrightarrow{\overline{\alpha}}} \hat{O}_s |\phi\rangle \langle \phi| \hat{O}_s^{\dag} \hat{M}_{\overrightarrow{\overline{\alpha}}}^{\dag}
\end{eqnarray}
where
\begin{eqnarray}
\hat{M}_{\overrightarrow{\overline{\alpha}}} = \langle \overrightarrow{\overline{\alpha}}| \hat{M}_{eve}| (\otimes_{l=1}^{\alpha -1} \hat{M}_l (\hat{O}_s |\phi\rangle)(\hat{O}_s |\phi\rangle)^{\otimes M_s-\alpha } (\hat{O}_f |\phi\rangle)^{\otimes M_f}  (|\phi\rangle)^{\otimes \overline{M}}.  
\end{eqnarray}
and the quantum states $\{ |\overrightarrow{\overline{\alpha}} \rangle \}$ represent the basis for the subsystems $A_s,B_s,A_c,B_c,A,B,a,b$ of all rounds in the protocol except the $\alpha$-th round in $\mathcal{M}_s$. 

Next, to derive Eq \eqref{eq:1}, we expand the quantum state $\hat{O}_s |\phi\rangle$ as 
\begin{eqnarray}
\hat{O}_s |\phi\rangle = p_{\mu} |00\rangle_{A_sB_s} |\phi\rangle_{\mu}+ p_O |11\rangle_{A_sB_s}|\phi\rangle_0 +  p_{\nu} |22\rangle_{A_sB_s} |\phi\rangle_{\nu}
\end{eqnarray}
where
\begin{eqnarray}
\label{eq:14}
&|\phi\rangle_{\mu} \\ \notag
=&  \sum_{l=0}^{M-1} \frac{1}{M} [ ( |ll\rangle_{A_cB_c}\frac{1}{2}(|0\rangle_A |e^{i\frac{2\pi}{M}l} \sqrt{\mu}\rangle_a + |1\rangle_A|-e^{i\frac{2\pi}{M}l}\sqrt{\mu} \rangle_a) ( |0\rangle_B |e^{i\frac{2\pi}{M}l} \sqrt{\mu}\rangle_b + |1\rangle_B|-e^{i\frac{2\pi}{M}l}\sqrt{\mu} \rangle_b) )  \\ \notag 
&+ (|l,(l+\frac{M}{2}) mod M\rangle_{A_c B_c}  \frac{1}{2} ( |0\rangle_A |e^{i\frac{2\pi}{M}l} \sqrt{\mu}\rangle_a + |1\rangle_A|-e^{i\frac{2\pi}{M}l}\sqrt{\mu} \rangle_a) ( |0\rangle_B |-e^{i\frac{2\pi}{M}l} \sqrt{\mu}\rangle_b + |1\rangle_B|e^{i\frac{2\pi}{M}l}\sqrt{\mu} \rangle_b))  )  ],   \notag 
\end{eqnarray}

\begin{eqnarray}
|\phi\rangle_{0}= |00\rangle_{A_cB_c}|00\rangle_{AB}|00\rangle_{ab}  
\end{eqnarray}
and 

\begin{small}
\begin{eqnarray}
&|\phi\rangle_{\nu}   \\  
=&  \sum_{l=0}^{M-1} \frac{1}{M} [ ( |ll\rangle_{A_cB_c}(|00\rangle_{AB} |e^{i\frac{2\pi}{M}l} \sqrt{\nu}\rangle_a   |e^{i\frac{2\pi}{M}l} \sqrt{\nu}\rangle_b  )   + (|l,(l+\frac{M}{2}) mod M\rangle_{A_c B_c} |00\rangle_{AB}  (  |e^{i\frac{2\pi}{M}l} \sqrt{\nu}\rangle_a  |-e^{i\frac{2\pi}{M}l} \sqrt{\nu}\rangle_b )  )  ].  \notag 
\end{eqnarray}
\end{small}
To summarize, each key bit can be viewed as an event that Eve announces a successful click conditioned that Alice and prepare $|\phi\rangle_{\mu}$ and measure $AB$ with $Z$-basis. Since the measurement on $AB$ made by Alice and Bob can be delayed after Eve's announcement of successful click, the phase error can be estimated by Alice and Bob measuring $AB$ with $X$-basis rather than $Z$-basis.  To get the phase error of this part, we rewrite $|\phi\rangle_{\mu}$ under $X$-bases of $AB$ as

\begin{eqnarray}
|\phi\rangle_{\mu} = \sum_{l=0}^{M-1} \frac{1}{M} (|ll\rangle_{A_cB_c} |\psi\rangle_{\mu,AaBb}^l+|l,(l+\frac{M}{2})modM\rangle_{A_cB_c} |\psi'\rangle_{\mu,AaBb}^l)
\label{eq:aa}
\end{eqnarray}
where
\begin{small}
\begin{align}
&|\psi\rangle_{\mu,AaBb}^l \notag \\
 =& \frac{1}{2} (|00\rangle_{AB} |e^{i\theta_l}\sqrt{\mu}\rangle_a |e^{i\theta_l}\sqrt{\mu}\rangle_b + |01\rangle_{AB}|e^{i\theta_l}\sqrt{\mu}\rangle_a |-e^{i\theta_l}\sqrt{\mu}\rangle_b  \notag \\
 &+ |10\rangle_{AB}|-e^{i\theta_l}\sqrt{\mu}\rangle_a |e^{i\theta_l}\sqrt{\mu}\rangle_b  
 + |11\rangle_{AB}|-e^{i\theta_l}\sqrt{\mu}\rangle_a |-e^{i\theta_l}\sqrt{\mu}\rangle_b)  \notag \\
=& \frac{1}{4} (|++\rangle_{AB}(|e^{i\theta_l}\sqrt{\mu}\rangle_a |e^{i\theta_l}\sqrt{\mu}\rangle_b + |e^{i\theta_l}\sqrt{\mu}\rangle_a |-e^{i\theta_l}\sqrt{\mu}\rangle_b + |-e^{i\theta_l}\sqrt{\mu}\rangle_a |e^{i\theta_l}\sqrt{\mu}\rangle_b + |-e^{i\theta_l}\sqrt{\mu}\rangle_a |-e^{i\theta_l}\sqrt{\mu}\rangle_b  )   \notag \\ 
&+ |--\rangle_{AB}( |e^{i\theta_l}\sqrt{\mu}\rangle_a |e^{i\theta_l}\sqrt{\mu}\rangle_b - |e^{i\theta_l}\sqrt{\mu}\rangle_a |-e^{i\theta_l}\sqrt{\mu}\rangle_b - |-e^{i\theta_l}\sqrt{\mu}\rangle_a |e^{i\theta_l}\sqrt{\mu}\rangle_b + |-e^{i\theta_l}\sqrt{\mu}\rangle_a |-e^{i\theta_l}\sqrt{\mu}\rangle_b  )  \notag \\
&+ |+-\rangle_{AB}(\cdots) + |-+\rangle_{AB}(\cdots)),
\end{align}
\end{small}

\begin{small}
\begin{align}
&|\psi'\rangle_{\mu,AaBb}^l \notag \\
=&  \frac{1}{2} (|00\rangle_{AB} |e^{i\theta_l}\sqrt{\mu}\rangle_a |-e^{i\theta_l}\sqrt{\mu}\rangle_b + |01\rangle_{AB}|e^{i\theta_l}\sqrt{\mu}\rangle_a |e^{i\theta_l}\sqrt{\mu}\rangle_b \notag  \\
&+ |10\rangle_{AB}|-e^{i\theta_l}\sqrt{\mu}\rangle_a |-e^{i\theta_l}\sqrt{\mu}\rangle_b + |11\rangle_{AB}|-e^{i\theta_l}\sqrt{\mu}\rangle_a |e^{i\theta_l}\sqrt{\mu}\rangle_b)  \notag \\
=&\frac{1}{4} (|++\rangle_{AB}(|e^{i\theta_l}\sqrt{\mu}\rangle_a |-e^{i\theta_l}\sqrt{\mu}\rangle_b + |e^{i\theta_l}\sqrt{\mu}\rangle_a |e^{i\theta_l}\sqrt{\mu}\rangle_b + |-e^{i\theta_l}\sqrt{\mu}\rangle_a |-e^{i\theta_l}\sqrt{\mu}\rangle_b + |-e^{i\theta_l}\sqrt{\mu}\rangle_a |e^{i\theta_l}\sqrt{\mu}\rangle_b  )   \notag \\ 
&+ |--\rangle_{AB}( |e^{i\theta_l}\sqrt{\mu}\rangle_a |-e^{i\theta_l}\sqrt{\mu}\rangle_b - |e^{i\theta_l}\sqrt{\mu}\rangle_a |e^{i\theta_l}\sqrt{\mu}\rangle_b - |-e^{i\theta_l}\sqrt{\mu}\rangle_a |-e^{i\theta_l}\sqrt{\mu}\rangle_b + |-e^{i\theta_l}\sqrt{\mu}\rangle_a |e^{i\theta_l}\sqrt{\mu}\rangle_b  )  \notag \\
&+ |+-\rangle_{AB}(\cdots) + |-+\rangle_{AB}(\cdots)),
\end{align}
\end{small}

and $\theta_l = \frac{l}{M}2\pi$. For the purpose of clarification, we define some  quantum states below: 
\begin{eqnarray}
|e^{i\theta}\sqrt{2\mu}\rangle_{ab} = \sum_{j=0 }^{\infty} e^{ij\theta}\sqrt{P_{j|2\mu}}|j\rangle_{ab}
\end{eqnarray}
and
\begin{align}
|e^{i\theta}\sqrt{2\mu}\rangle_{ab,even}  = \frac{|e^{i\theta}\sqrt{\mu}\rangle_a|e^{i\theta}\sqrt{\mu}\rangle_b+ |-e^{i\theta}\sqrt{\mu}\rangle_a|-e^{i\theta}\sqrt{\mu}\rangle_b}{2}  = \sum_{j \in \mathcal{N}_0} e^{ij\theta}\sqrt{P_{j|2\mu}}|j\rangle_{ab},
\end{align}
where $|j\rangle_{ab} = \sum_{i=0}^{j}\sqrt{\frac{j!}{2^j i!(j-i)!} }|i\rangle_a|j-i\rangle_b$ and $\mathcal{N}_0$ is the set of even numbers. Indeed, $|j\rangle$ is a quantum state satisfying that the total photon-number of $a$ and $b$ is $j$. Besides, another similar quantum state is defined below:

\begin{eqnarray}
|e^{i\theta}\sqrt{2\mu}\rangle_{ab}' = \sum_{j=0 }^{\infty} e^{ij\theta}\sqrt{P_{j|2\mu}}|j\rangle_{ab}'
\end{eqnarray}
and 
\begin{align}
|e^{i\theta}\sqrt{2\mu}\rangle_{ab}'  = \frac{|e^{i\theta}\sqrt{\mu}\rangle_a|-e^{i\theta}\sqrt{\mu}\rangle_b+ |-e^{i\theta}\sqrt{\mu}\rangle_a|e^{i\theta}\sqrt{\mu}\rangle_b}{2}  = \sum_{j \in \mathcal{N}_0} e^{ij\theta} \sqrt{P_{j|2\mu}}|j\rangle'_{ab}
\end{align}
where $|j\rangle'_{ab,even} = \sum_{i=0}^{j}(-1)^i\sqrt{\frac{j!}{2^{j} i!(j-i)!} }|i\rangle_a|j-i\rangle_b$.
 With these definitions, we can write the quantum states   $|\psi\rangle_{\mu,AaBb}^l$ and $ |\psi'\rangle_{\mu,AaBb}^l $  in a more simplified way, namely
 \begin{align}
 |\psi\rangle_{\mu,AaBb}^l 
 =&\frac{1}{2} ( |++\rangle_{AB}( |e^{i\theta_l}\sqrt{2\mu}\rangle_{ab,even} +  |e^{i\theta_l}\sqrt{2\mu}\rangle_{ab,even}' ) +|--\rangle_{AB}( |e^{i\theta_l}\sqrt{2\mu}\rangle_{ab,even} - |e^{i\theta_l}\sqrt{2\mu}\rangle_{ab,even}' )   \notag \\
 &+ |+-\rangle_{AB}(\cdots) + |-+\rangle_{AB}(\cdots) )
 \end{align}
 and
  \begin{align}
 |\psi'\rangle_{\mu,AaBb}^l 
 =&\frac{1}{2} ( |++\rangle_{AB}( |e^{i\theta_l}\sqrt{2\mu}\rangle_{ab,even} +  |e^{i\theta_l}\sqrt{2\mu}\rangle_{ab,even}' ) -|--\rangle_{AB}( |e^{i\theta_l}\sqrt{2\mu}\rangle_{ab,even} - |e^{i\theta_l}\sqrt{2\mu}\rangle_{ab,even}' )   \notag \\
 &+ |+-\rangle_{AB}(\cdots) + |-+\rangle_{AB}(\cdots) )
 \end{align}
 Obviously, the measurement outcome of $|++\rangle_{AB}$ and $|--\rangle_{AB}$ can be defined as phase error event. Recall the whole density matrix  $|\phi\rangle_{\mu}\langle \phi| $ given in Eq \eqref{eq:14}, we can get the $ab$ part corresponding to the phase error event, 
 \begin{eqnarray}
 \hat{\rho}_{\mu,ph} =& \langle ++|_{AB}tr_{A_cB_c}(|\phi\rangle_{\mu}\langle \phi|)|++\rangle_{AB} +\langle --|_{AB}tr_{A_cB_c}(|\phi\rangle_{\mu}\langle \phi|)|--\rangle_{AB} \notag \\
 =& \sum_{l=0}^{M-1} \langle ll|_{A_cB_c} (\langle ++|_{AB} + \langle--|_{AB}) |\phi\rangle_{\mu}\langle \phi| (|++\rangle_{AB} + |--\rangle_{AB }) |ll\rangle_{A_cB_c}  \notag \\
 & +\langle l,(l+M/2)modM|_{A_cB_c} (\langle ++|_{AB} + \langle--|_{AB}) |\phi\rangle_{\mu}\langle \phi| (|++\rangle_{AB} + |--\rangle_{AB }) |l,(l+M/2)modM\rangle_{A_cB_c}  \notag \\
 =& \sum_{l=0}^{M-1} \frac{1}{ M^2} (|e^{i\theta_l}\sqrt{2\mu} \rangle_{ab}\langle e^{i\theta_l}\sqrt{2\mu}|  + |e^{i\theta_l}\sqrt{2\mu} \rangle_{ab}'\langle e^{i\theta_l}\sqrt{2\mu}| )
 \end{eqnarray}
 According to Eq(2.5)- Eq(2.7) of Ref \cite{cao2015discrete}, $\hat{\rho}_{\mu,ph}$ can be rewritten as
 \begin{eqnarray}
 \hat{\rho}_{\mu,ph} =&  \frac{2}{M} \sum_{j=0,j\in \mathcal{N}_0}^{M-2} \tilde{P}_{j|2\mu}(\frac{1}{2} |\tilde{j}_{2\mu}\rangle_{ab} \langle \tilde{j}_{2\mu}| + \frac{1}{2} |\tilde{j}_{2\mu}\rangle'_{ab} \langle \tilde{j}_{2\mu}| ) \notag \\
 =&\frac{2}{M} \sum_{j=0,j\in \mathcal{N}_0}^{M-2} \tilde{P}_{j|2\mu}\tau_{j|2\mu},
 \end{eqnarray}
 where $ \tilde{P}_{j|2\mu}  = \sum_{n=0}^{\infty} P_{j+Mn|2\mu}$, $P_{j|2\mu}$ is the the probability of finding $j$ photons in a Poisson source with mean photon-number $2\mu$, $|\tilde{j}_{2\mu}\rangle = \sum_{n=0}^{\infty} \frac{\sqrt{P_{j+Mn|2\mu}} }{\sqrt{ \tilde{P}_{j|2\mu}} } |j+Mn\rangle_{ab} $ and $|\tilde{j}_{2\mu}\rangle' = \sum_{n=0}^{\infty} \frac{\sqrt{P_{j+Mn|2\mu}} }{\sqrt{ \tilde{P}_{j|2\mu}} } |j+Mn\rangle_{ab}'$, and $\tau_{j|2\mu}=\frac{1}{2} |\tilde{j}_{2\mu}\rangle_{ab} \langle \tilde{j}_{2\mu}| + \frac{1}{2} |\tilde{j}_{2\mu}\rangle'_{ab} \langle \tilde{j}_{2\mu}|$. 
 
 For ease of understanding, we can interpret the formula of $\hat{\rho}_{\mu,ph}$ in an easy way.
 It's easy to see that $\hat{\rho}_{\mu,ph}$ is a mixture of $\tau_{j|2\mu}$, which consists of photon-number state $|j+Mn\rangle_{ab},n=0,1,2,...$, and the probability of finding $j+Mn$ photons is proportional to $P_{j+Mn|2\mu}$. Let $\tau_{even|2\mu}$ be the normalized $\hat{\rho}_{\mu,ph}$, then a phase error event for a key bit is equivalent to a successful click announced by Eve conditioned on that a mixture $\tau_{even|2\mu}=\sum_{j=0,j\in \mathcal{N}_0}^{M-2}\tilde{P}_{j|2\mu}\tau_{j|2\mu}/P_{even|2\mu}$  prepared by Alice and Bob, and the probability of preparing such a mixture is obviously $P^2_\mu P_{even|2\mu}2/M$.

To find a way to estimate the number of phase error, we can give the density matrices Alice and Bob prepared in code mode and decoy mode. If we trace out $ABA_cB_c$ of the quantum state $|\phi\rangle_{\mu}\langle\phi|$, i.e. no matter the measurement outcome on $AB$ is phase error or not,  we have 
  
\begin{eqnarray}
\hat{\rho}_{\mu} =& Tr_{ABA_cB_c}(|\phi\rangle_{\mu}\langle\phi|)  \notag  \\ 
=& \frac{2}{M} \sum_{j=0}^{M-1} \tilde{P}_{j|2\mu}(  \frac{1}{2} |\tilde{j}_{2\mu} \rangle \langle \tilde{j}_{2\mu} | + \frac{1}{2} |\tilde{j}_{2\mu} \rangle' \langle \tilde{j}_{2\mu} | ) \notag \\
=& \frac{2}{M} \sum_{j=0}^{M-1} \tilde{P}_{j|2\mu} \tau_{j|2\mu}.
\end{eqnarray}
We define $\tau_{2\mu}=\sum_{j=0}^{M-1}\tilde{P}_{j|2\mu}\tau_{j|2\mu}$ is the normalized $\hat{\rho}_{\mu}$. The genaration of a key bit is equivalent to a successful click announced by Eve conditioned on that a mixture $\tau_{2\mu}=\sum_{j=0}^{M-1}\tilde{P}_{j|2\mu}\tau_{j|2\mu}$ is prepared by Alice and Bob, and the probability of preparing such a mixture is obviously $P^2_\mu 2/M$.

Now we are approaching a main result of above derivations. In the $N_{tot}$ rounds, Alice and Bob prepare $\tau_{2\mu}$ with probability $P^2_\mu 2/M$, the number of its successful rounds is denoted by $n_{2\mu}$. And of course, the key bits are generated in these rounds, thus $n_{2\mu}=n_{bit}$. Since $\tau_{2\mu}$ is a mixture of $ \tau_{j|2\mu},j=0,1...,,M-1$,  the $n_{2\mu}$ successful clicks are sum of clicks by $\tau_{j|2\mu},j=0,1,...,M-1$.  Then $n_{2\mu}=\sum_{j=0}^{M-1}n_{j|2\mu}$ holds evidently, in which $n_{j|2\mu}$ is the number of successful clicks by $\tau_{j|2\mu}$. Recall that the phase error for these events are from clicks by $\tau_{even|2\mu}$, one can assert that the number of phase error events $n_{ph}=\sum_{j=0,j\in \mathcal{N}_0}^{M-2}n_{j|2\mu}$ must hold. This is a main result we have so far.

\subsection{Linear programming for estimating the number of phase error events}

We have proved that $n_{ph}=\sum_{j=0,j\in \mathcal{N}_0}^{M-2}n_{j|2\mu}$  with the  constraint $n_{2\mu}=\sum_{j=0}^{M-1}n_{j|2\mu}$. Obviously, this is not sufficient for estimating $n_{ph}$ tightly. Here, we resort to decoy states to get more constraints to bound $n_{ph}$

Similarly with the analysis of clicks by $\hat{\rho}_{\mu} =Tr_{ABA_cB_c}(|\phi\rangle_{\mu}\langle\phi|) $,  a successful click from decoy mode with intensity $\nu$ means that Alice and Bob prepare  

\begin{eqnarray}
\hat{\rho}_{\nu} =& Tr_{ABA_cB_c}(|\phi\rangle_{\nu}\langle\phi|)  \notag  \\ 
=& \frac{2}{M} \sum_{j=0}^{M-1} \tilde{P}_{j|2\nu}(  \frac{1}{2} |\tilde{j}_{2\nu} \rangle \langle \tilde{j}_{2\nu} | + \frac{1}{2} |\tilde{j}_{2\nu} \rangle' \langle \tilde{j}_{2\nu} | ) \notag \\
=& \frac{2}{M} \sum_{j=0}^{M-1} \tilde{P}_{j|2\nu} \tau_{j|2\nu}.
\end{eqnarray}
Accordingly,  we also have $n_{2\nu}= \sum_{j=0}^{M-1} n_{j|2\nu}$. Here, $n_{2\nu}$ and $n_{j|2\nu}$ are defined analogously to $n_{2\mu}$ and $n_{j|2\mu}$ respectively. Intuitively, $n_{j|2\mu}$ and $n_{j|2\mu}$ are the numbers of sucessful clicks for $\tau_{j|2\mu}$ and $\tau_{j|2\nu}$ respectively. Typically $\mu<\nu<<1$ is satisfied, then both $\tau_{j|2\mu}$ and $\tau_{j|2\nu}$ are very close to the photon-number state $|j\rangle_{ab}$. This implies that the gap between $n_{j|2\mu}$ and $n_{j|2\nu}$ can be bounded, and then we may estimate $n_{ph}=\sum_{j=0,j\in \mathcal{N}_0}^{M-2}n_{j|2\mu}$. Indeed, with the result in appendix B of \cite{cao2015discrete}, the gap between $n_{j|2\mu}$ and $n_{j|2\nu}$ in asymptotic case can be obtained. Here, we develop Lemma1 to bound this gap in finite-key situations.

\begin{lemma}
	If Alice prepares $N_{tot}$ pairs of particles $A$ and $B$ with the quantum state $(\sum_{i} \sqrt{P_i}|i\rangle_A |\phi_i\rangle_B)^{\otimes N_{tot}}$ where $\langle i | j \rangle  = \delta_{ij},|\langle \phi_i |\phi_j \rangle| = F_{ij}$ and she sends the B part in each pair to Eve. For every round, Eve announces the measurement is successful or unsuccessful which is denoted by $M=1$ or $M=0$ respectively. Then Alice measures the subsystem A with projectors $\{|i\rangle\langle i|,i=0,1,2,...\}$ to which quantum state she sent for the pairs that Eve announced $M=1$. Let $n_i$ denote the number of yields for the quantum state $|i\rangle$. If $P_i \textgreater P_j$, we have  the constraints between $n_i$ and $n_j$, say, 
	\begin{eqnarray}
		|\frac{P_j}{P_i}n_i - n_j| \le N_1   \sqrt{1-F_{ij}^2 } + 2\delta( N_1,\frac{1+\sqrt{ 1-F^2_{ij}}}{2} ),\varepsilon_0) -2 \delta(N_2-n_i-n_j,2P_j,\varepsilon_0) + \delta(n_i,\frac{P_j}{P_i},\varepsilon_2) 
		\label{eq:27}
	\end{eqnarray}
	holds with a failure probability $2\varepsilon_0+2\varepsilon_1+2\varepsilon_2$, where
	\begin{eqnarray}
		\delta(x,y,z)  =& \sqrt{3xy\ln(\frac{1}{z})} \notag \\ 
		N_1=&2P_jN_{tot} +\delta(N_{tot},2P_j,\varepsilon_1) \notag \\
		N_2=&2P_jN_{tot} -\delta(N_{tot},2P_j,\varepsilon_1)
	\end{eqnarray}
\end{lemma}

\begin{proof}
	Since we are only interested in the statistics of $n_i$ and $n_j$, it's not restrictive to rewrite the quantum state $(\sum_{i} \sqrt{P_i}|i\rangle_A |\phi_i\rangle_B)^{\otimes N_{tot}}$ as 
	\begin{eqnarray}
		(\sum_{i} \sqrt{P_i}|i\rangle_A |\phi_i\rangle_B)^{\otimes N_{tot}} = \{ \sqrt{P_i-P_j} |i'\rangle_A |\phi_i\rangle_B + \sqrt{2P_j} \frac{1}{\sqrt{2}} (|i^{''}\rangle_A|\phi_i\rangle_B + |j\rangle_A |\phi_j\rangle_B)+ ... \}^{\otimes N_{tot}}. 
	\end{eqnarray}
	Here, we virtually define $\sqrt{P_i}|i\rangle_A=\sqrt{P_i-P_j}|i'\rangle_A+\sqrt{P_j}|i''\rangle_A$ and $\langle i'|i''\rangle_A=0$, which do not change the density matrix of $B$, thus have no impact on Eve's operation and statistics of $n_i$ and $n_j$. 
	Let $n_{i'}$($n_{i''}$) denote the number of detections for the quantum  state $|i'\rangle_A $($ |i''\rangle_A$) when Eve announces a successful measurement. Apparently, we know that $n_i = n_{i'}+ n_{i''}$. 
	
	Let's focus on the state $ \{ \sqrt{2P_j} \frac{1}{\sqrt{2}} (|i^{''}\rangle_A|\phi_i\rangle_B + |j\rangle_A |\phi_j\rangle_B) \}^{\otimes N_{tot}} $, by which the relation between $n_{i''}$ and $n_j$ can be analyzed. The essential idea is reinterpreting the detection of $B$ to a game of Eve guessing which states $|\phi_i\rangle_B$ or $|\phi_j\rangle_B$ Alice prepared for all of the $N_{tot}$ trials. Specifically, we consider a virtual experiment illustrated below.
	
	Alice prepares the quantum state $\{ \sqrt{2P_j} \frac{1}{\sqrt{2}} (|i^{''}\rangle_A|\phi_i\rangle_B + |j\rangle_A |\phi_j\rangle_B) \}^{\otimes N_{tot}}$, then she sends the $B$ part to Eve. Eve measures each $B$ she received. If she gets a successful measurement, she will announce $M =1$. Otherwise, she will announce $M=0$. Up to now, there's no difference from previous protocol. A critical step is that for any trial that Eve announces $M=0$, Alice flips corresponding $M$ with probability $\frac{1}{2}$. Finally, Alice measures all partials $A$ locally. Without imposing any limitation on Eve, we can reinterpret that  $M=1$ ($M=0$) means that Eve guessed the quantum state Alice prepared is $|\phi_i\rangle $ ($|\phi_j\rangle $). In a word, we can now treat this virtual experiment as a game where Eve tries to guess Alice preparing $|\phi_i\rangle$ or $|\phi_j\rangle $. For each of all the trials in such a game, it is well known that Eve's maximal probability of guessing correctly is $\frac{1+\sqrt{1-F_{ij}^2}}{2}$. Now, we are ready to find the relation between $n_{i''}$ and $n_j$ by calculating how many trials in which Eve's guessing is correct. First, $n_{i''}$ means that announcing $M=1$ at first and Alice also preparing $|\phi_i\rangle_B$, which of course leads to guessing correctly. Then let $N_{i''} $($N_j $) denote Alice preparing the state $ |\phi_i\rangle $($|\phi_j\rangle $), which implies that $N_{i''}+N_j\approx N_{tot}2P_j$. As a result, there are $(N_{i''}+N_{j}-n_{i''}-n_{j})$ trials in which Eve announcing $M=0$ at first and then a random flipping operation on $M$ is following; for each of such trials, the probability of guessing correctly is obviously $1/2$. Further considering the potential statistical fluctuations made by the random flipping, with a failure probability of $\varepsilon_1$, the number of Eve guessing correctly in the $N_{i''}+N_j$ trials is no larger than 
	\begin{eqnarray}
		n_{i''} + \frac{1}{2}(N_{i''}-n_{i''}) + \frac{1}{2}(N_j-n_j)  + \overline{\delta_1},
	\end{eqnarray}
	where $\overline{\delta_1} =\delta(N_{i''}+N_j-n_{i''}-n_j,\frac{1}{2},\varepsilon_1)$ is the upper bound of the  statistical fluctuation made by the random flipping of the $(N_{i''}+N_{j}-n_{i''}-n_{j})$ trials. On the other hand, in the $N_{i''}+N_j$ trials of guessing $|\phi_i\rangle$ and $|\phi_j\rangle $ prepared by Alice at random, the probability of Eve guessing correctly is no larger than $\frac{1+\sqrt{1-F_{ij}^2}}{2}$\cite{ivanovic1987differentiate}, because the fidelity of $|\phi_i\rangle$ and $|\phi_j\rangle $ is $F_{ij}$,  Hence, one can assert that with a failure probability $\varepsilon_0$,
	\begin{eqnarray}
		\frac{n_{i''}-n_j}{2} + \frac{N_{i''}+N_j}{2} + \overline{\delta_1} \leq (N_{i''}+N_j) \frac{1+\sqrt{1-F_{ij}^2 }}{2}+ \overline{\delta_2}
		\label{eq:31}
	\end{eqnarray}
	holds, where $\overline{\delta_2} = \delta(N_{i''}+N_j,\frac{1+\sqrt{1-F_{ij}^2 }}{2},\varepsilon_0)$ is the upper bound of the statistical fluctuation when Eve's guessing probability for each trial achieves the upper-bound $\frac{1+\sqrt{1-F_{ij}^2}}{2}$.
	
	According to Eq \eqref{eq:31}   we can get an upper bound of $n_{i''}-n_j$ with a failure probability $\varepsilon_1$, say 
	\begin{eqnarray}
		n_{i''}-n_j \le ( N_{i''}+N_j)\sqrt{1-F_{ij}^2}  +2\delta(N_{i''}+N_j,\frac{1+\sqrt{1-F_{ij}^2 }}{2},\varepsilon_0) -2\delta(N_{i''}+N_j-n_{i''}-n_j,\frac{1}{2},\varepsilon_0).
		\label{eq:39}
	\end{eqnarray}
	Similarly, if we redefine the guessing correctly as $M=1$ corresponding to $|\phi_j\rangle$ and $M=0$ corresponding to $|\phi_i\rangle$,  we have that 
	\begin{eqnarray}
		n_{j}-n_i'' \le ( N_{i''}+N_j)\sqrt{1-F_{ij}^2}  +2\delta(N_{i''}+N_j,\frac{1+\sqrt{1-F_{ij}^2 }}{2},\varepsilon_0) -2\delta(N_{i''}+N_j-n_{i''}-n_j,\frac{1}{2},\varepsilon_0)
		\label{eq:40}
	\end{eqnarray}
	Combining Eq \eqref{eq:39} and Eq \eqref{eq:40}, we are clear that 
	\begin{eqnarray}
		|n''_{i}-n_j| \le ( N_{i''}+N_j)\sqrt{1-F_{ij}^2}  +2\delta(N_{i''}+N_j,\frac{1+\sqrt{1-F_{ij}^2 }}{2},\varepsilon_0) -2\delta(N_{i''}+N_j-n_{i''}-n_j,\frac{1}{2},\varepsilon_0)
		\label{eq:41}
	\end{eqnarray}
	holds with a failure probability $2\varepsilon_0$.
	
	For simplicity, we enlarge the R.H.S of Eq \eqref{eq:41}. Concretely, we replace $ 2\delta(N_{i''}+N_j-n_{i''}-n_j,\frac{1}{2},\varepsilon_0)$ by $ 2\delta(N_{i''}+N_j-n_i-n_j,\frac{1}{2},\varepsilon_0)$ in Eq \eqref{eq:41}, say 
	\begin{eqnarray}
		|n_{i''}-n_j| \le ( N_{i''}+N_j)\sqrt{1-F_{ij}^2}  +2\delta(N_{i''}+N_j,\frac{1+\sqrt{1-F_{ij}^2 }}{2},\varepsilon_0) -2\delta(N_{i''}+N_j-n_i-n_j,\frac{1}{2},\varepsilon_0).
		\label{eq:42}
	\end{eqnarray}
	Note that these bounds of statistical fluctuations can be derived by the Chernoff bound \cite{zhang2017improved}.
	
	Similarly, because the probability that Alice sends the quantum state $\frac{1}{\sqrt{2}} (|i^{''}\rangle_A|\phi_i\rangle_B + |j\rangle_A |\phi_j\rangle_B)$   is $2P_j$, using the well-known Chernoff bound \cite{zhang2017improved}, we know that 
	
	\begin{eqnarray}
		2N_{tot}P_j-\delta(N_{tot},2P_j,\varepsilon_1)  \le N_{i''} + N_j \le 2N_{tot}P_j +  \delta(N_{tot},2P_j,\varepsilon_1)
		\label{eq:43}
	\end{eqnarray}
	Combing Eq \eqref{eq:42} and Eq \eqref{eq:43}, we know that 
	\begin{eqnarray}
		|n_{i''}-n_j| \le N_1\sqrt{1-F_{ij}^2}  +2\delta(N_1,\frac{1+\sqrt{1-F_{ij}^2 }}{2},\varepsilon_0) -2\delta(N_2-n_i-n_j,\frac{1}{2},\varepsilon_0)
		\label{eq:44}
	\end{eqnarray}
	holds with a failure probability $2\varepsilon_0+2\varepsilon_1$, 
	where $N_1 = 2N_{tot}P_j+\delta(N_{tot},2P_j,\varepsilon_1)$ and $N_2 = 2N_{tot}P_j -  \delta(N_{tot},2P_j,\varepsilon_1)$

	Finally, to  derive the relation between $n_i$ and $n_j$, we have to consider the relations between $n_i$and $n_{i''}$. It is easy to know that $n_{i''} = \frac{P_j}{P_i} n_i$ on average, since there is no way for Eve to distinguish $|i'\rangle$ and $|i''\rangle$. Hence, using the Chernoff bound \cite{zhang2017improved} again, we know that
	\begin{eqnarray}
		\frac{P_j}{P_i}n_i - \delta(n_i,\frac{P_j}{P_i},\varepsilon_2) \le n_{i''} \le \frac{P_j}{P_i}n_i + \delta(n_i,\frac{P_j}{P_i},\varepsilon_2)
		\label{eq:45}
	\end{eqnarray}  
	Combining Eq \eqref{eq:44} and Eq \eqref{eq:45}, we can get the inequality Eq \eqref{eq:27}. 
	
	In conclude, we complete the proof.
\end{proof}
 With Lemma1, we can derive the constrains between $n_{i|2\mu}$ and $n_{j|2\nu}$. Since Alice and Bob send the quantum state $\tau_{j|2\mu} $   ($ \tau_{j|2\nu}$ ) with the probability $ 
 \frac{2P_{\mu}^2 }{M} \tilde{P}_{j|2\mu}$($\frac{2P_{\nu}^2 }{M} \tilde{P}_{j|2\nu} $) and the fidelity between them is $F^j_{\mu\nu} = \sum_{n=0}^{\infty}\frac{\sqrt{P_{j+Mn|2\mu}} \sqrt{P_{j+Mn|2\nu}} }{\sqrt{\tilde{P}_{j|2\mu} \tilde{P}_{j|2\nu}  }}$. By Lemma1, we can get the relation between $n_{j|2\mu}$ and $n_{j|2\nu}$ which reads
 \begin{eqnarray}
 	|C_{j|2\mu} n_{j|2\mu} - C_{j|2\nu} n_{j|2\nu}| \le \triangle^j_{\mu\nu}. 
 	\label{eq:46}
 \end{eqnarray}
 
 We let $P_1 = \frac{2P_{\mu}^2 }{M} \tilde{P}_{j|2\mu} $ and $ P_2 = \frac{2P_{\nu}^2 }{M} \tilde{P}_{j|2\nu} $. 
 If $P_1 > P_2$, we have $C_{j|2\mu} = \frac{P_2}{P_1}$,$C_{j|2\nu} =1$ and 
 \begin{eqnarray}
 	\triangle^j_{\mu\nu} = N_1\sqrt{1-(F_{\mu\nu}^j)^2}  +2\delta(N_1,\frac{1+\sqrt{1-(F_{\mu\nu}^j)^2 }}{2},\varepsilon_0) -2\delta(N_2-n_{j|2\mu}-n_{j|2\nu},\frac{1}{2},\varepsilon_0)+ \delta(n_{j|2\mu},\frac{P_2  }{P_1},\varepsilon_2)  
 	\label{eq:47}
 \end{eqnarray} 
 where $N_1 =2N_{tot}P_2 + \delta(N_{tot},2P_2,\varepsilon_1) $ and $ N_2 = 2N_{tot}P_2 - \delta(N_{tot},2P_2,\varepsilon_1)$
 
 If $ P_1 < P_2$,  we have that $C_{j|2\nu} = \frac{P_1  }{P_2 }$,$C_{j|2\mu} =1$ and 
 \begin{eqnarray}
 	\triangle^j_{\mu\nu}  = N_1\sqrt{1-(F_{\mu\nu}^j)^2}  +2\delta(N_1,\frac{1+\sqrt{1-(F_{\mu\nu}^j)^2 }}{2},\varepsilon_0) -2\delta(N_2-n_{j|2\mu}-n_{j|2\nu},\frac{1}{2},\varepsilon_0)+\delta(n_{j|2\nu},\frac{P_1  }{P_2},\varepsilon_2)  
 	\label{eq:48}
 \end{eqnarray}
 where $N_1 =2N_{tot}P_1 + \delta(N_{tot},2P_1,\varepsilon_1) $ and $ N_2 = 2N_{tot}P_1 - \delta(N_{tot},2P_1,\varepsilon_1)$
 
 One can know that the constraints Eq \eqref{linearprogramming} and Eq \eqref{eq:48} are nonlinear because of the $n_{j|2\mu} $ in $ \delta(n_{j|2\mu},\frac{P_2}{P_1},\varepsilon_2) $ or $n_{j|2\nu} $ in $ \delta(n_{j|2\nu},\frac{P_1}{P_2},\varepsilon_2) $. To keep the linearity of these constraints for ease of numerical calculations, we replace $n_{j|2\mu}$ with $ n_{2\mu}$ and replace
 $n_{j|2\nu}$ with $ n_{2\nu}$ in $\triangle^j_{\mu\nu}$. That's to say, without compromising the security, we replace Eq \eqref{eq:47} by 
 \begin{eqnarray}
 	\triangle^j_{\mu\nu} = N_1\sqrt{1-(F_{\mu\nu}^j)^2}  +2\delta(N_1,\frac{1+\sqrt{1-(F_{\mu\nu}^j)^2 }}{2},\varepsilon_0) -2\delta(N_2-n_{2\mu}-n_{2\nu},\frac{1}{2},\varepsilon_0)+ \delta(n_{2\mu},\frac{P_2  }{P_1},\varepsilon_2)
 \end{eqnarray}
 and replace Eq \eqref{eq:48} by
 \begin{eqnarray}
 	\triangle^j_{\mu\nu}  = N_1\sqrt{1-(F_{\mu\nu}^j)^2}  +2\delta(N_1,\frac{1+\sqrt{1-(F_{\mu\nu}^j)^2 }}{2},\varepsilon_0) -2\delta(N_2-n_{2\mu}-n_{2\nu},\frac{1}{2},\varepsilon_0)+\delta(n_{2\nu},\frac{P_1  }{P_2},\varepsilon_2)  
 \end{eqnarray}
 
 Besides, recalling Alice and Bob may both choose intensity $0$ and obtain the corresponding number of successful clicks $n_0$,  we have two additional constraints between $n_0$ and $n_{0|2\mu},n_{0|2\nu}$ which reads
 \begin{eqnarray}
 	|C_{0,\mu} n_0 - C^0_{2\mu} n_{0|2\mu}| \le  \triangle^0_{0\mu} \notag \\
 	|C_{0,\nu} n_0 - C^0_{2\nu} n_{0|2\nu}| \le  \triangle^0_{0\nu}
 	\label{eq:51}
 \end{eqnarray}
 Since  the probability of both Alice and Bob send the quantum state $|0\rangle\langle 0|$ is $P_O^2$ and the fidelity between $|0\rangle\langle 0|$ and $\frac{1}{2} |\tilde{0}_{2\mu} \rangle \langle \tilde{0}_{2\mu} | + \frac{1}{2} |\tilde{0}_{2\mu} \rangle' \langle \tilde{0}_{2\mu} | $   ($ \frac{1}{2} |\tilde{0}_{2\nu} \rangle \langle \tilde{0}_{2\nu} | + \frac{1}{2} |\tilde{0}_{2\nu} \rangle' \langle \tilde{0}_{2\nu} |$ ) is $ F^0_{\mu0}= \frac{ P_{0|2\mu} }{\tilde{P}_{0|2\mu} } $($ F^0_{\nu0} =\frac{ P_{0|2\nu} }{\tilde{P}_{0|2\nu} }$),  we can get the coefficients of these two constraints according to Lemma1. For simplicity we let $P_1 =P_O^2 $ and $P_2 = \frac{2 P_{\mu}^2 \tilde{P}_{0|2\mu}}{M}$, then if $P_1 > P_2$, we have 
 $C_{0,\mu}  = \frac{P_2}{P_1},C_{2\mu}^0 = 1 $ and
 \begin{eqnarray}
 	\triangle^0_{0\mu} =  N_1\sqrt{1-(F_{\mu 0}^0)^2}  +2\delta(N_1,\frac{1+\sqrt{1-(F_{\mu 0}^0)^2 }}{2},\varepsilon_0) -2\delta(N_2-n_{2\mu}-n_{0},\frac{1}{2},\varepsilon_0) + \delta(n_0,\frac{P_2}{P_1},\varepsilon_2) \notag 
 \end{eqnarray}
 where $N_1 = 2N_{tot}P_2 + \delta({N_{tot},2P_2,\varepsilon_1}) $ and $ N_2 = 2N_{tot}P_2 - \delta({N_{tot},2P_2,\varepsilon_1}) $;
 otherwise, we have $ C_{0,\mu} = 1,C_{2\mu}^0 =\frac{P_1}{P_2}$
 and $ C_{0,\mu}  = \frac{P_2}{P_1},C_{2\mu}^0 = 1 $ and
 \begin{eqnarray}
 	\triangle^0_{0\mu} =  N_1\sqrt{1-(F_{\mu 0}^0)^2}  +2\delta(N_1,\frac{1+\sqrt{1-(F_{\mu 0}^0)^2 }}{2},\varepsilon_0) -2\delta(N_2-n_{2\mu}-n_{0},\frac{1}{2},\varepsilon_0)+\delta(n_{2\mu},\frac{P_1}{P_2},\varepsilon_2)  
 \end{eqnarray}
 where $N_1 = 2N_{tot}P_1 + \delta({N_{tot},2P_1,\varepsilon_1}) $ and $ N_2 = 2N_{tot}P_1 - \delta({N_{tot},2P_1,\varepsilon_1}) $.
 
 Similarly,  we let $P_1 = P_O^2$ and $P_2 = \frac{2P_{\nu}^2 \tilde{P}_{0|2\nu}}{M} $, then if $P_1 > P_2$, we have 
 $C_{0,\nu}  = \frac{P_2}{P_1},C_{2\nu}^0 = 1 $ and
 \begin{eqnarray}
\triangle^0_{0\nu} =  N_1\sqrt{1-(F_{\nu 0}^0)^2}  +2\delta(N_1,\frac{1+\sqrt{1-(F_{\nu 	0}^0)^2 }}{2},\varepsilon_0) -2\delta(N_2-n_{2\nu}-n_0,\frac{1}{2},\varepsilon_0) +  
\delta(n_0,\frac{P_2}{P_1},\varepsilon_2)\notag
\end{eqnarray}
 where $N_1 = 2N_{tot}P_2 + \delta({N_{tot},2P_2,\varepsilon_1}) $ and $ N_2 = 2N_{tot}P_2 - \delta({N_{tot},2P_2,\varepsilon_1}) $;
 otherwise, we have $ C_{0,\nu} = 1, C_{2\nu}^0 =\frac{P_1}{P_2}$
 and $ C_{0,\nu}  = \frac{P_2}{P_1},C_{2\nu}^0 = 1 $ and
 \begin{eqnarray}
 	\triangle^0_{0\nu} =  N_1\sqrt{1-(F_{\nu 0}^0)^2}  +2\delta(N_1,\frac{1+\sqrt{1-(F_{\nu 0}^0)^2 }}{2},\varepsilon_0) -2\delta(N_2-n_{2\nu}-n_{0},\frac{1}{2},\varepsilon_0)+\delta(n_{2\nu},\frac{P_1}{P_2},\varepsilon_2)  
 \end{eqnarray}
 where $N_1 = 2N_{tot}P_1 + \delta({N_{tot},2P_1,\varepsilon_1}) $ and $ N_2 = 2N_{tot}P_1 - \delta({N_{tot},2P_1,\varepsilon_1}) $.
 
 Now the gaps between $n_{j|2\mu}$ v.s. $n_{j|2\nu}$, $n_{0|2\mu}$ v.s. $n_{0}$, and $n_{0|2\nu}$ v.s. $n_{0}$ have been given. To bound $n_{ph}$, we can now resort to a linear programming below:
 
 \begin{small}
 	\begin{alignat}{2}
 	\max \quad & n_{ph}=  \sum_{j=0,j\in \mathcal{N}_0}^{M-2}n_{j|2\mu}   \notag \\
 	\mbox{s.t.} \quad  
 	& \sum_{j=0}^{M-1} n_{j|2\mu}  =  n_{2\mu}  \notag \\
 	& \sum_{j=0}^{M-1} n_{j|2\nu} =  n_{2\nu} \notag  \\
 	& |C_{0,\mu} n_0 - C^0_{2\mu} n_{0|2\mu}| \le  \triangle^0_{0\mu} \notag \\
 	& |C_{0,\nu} n_0 - C^0_{2\nu} n_{0|2\nu}| \le  \triangle^0_{0\nu} \notag \\
 	& |C_{j|2\mu}n_{j|2\mu}  - C_{j|2\nu} n_{j|2\nu}| \le \triangle^j_{\mu\nu}  \notag \\
 	& 0 \le n_{j|2\mu} \le  N_{tot}\frac{2P_{\mu}^2 }{M} \tilde{P}_{j|2\mu} + \sqrt{3 \ln(1/\varepsilon_a)N_{tot}\frac{2P_{\mu}^2 }{M} \tilde{P}_{j|2\mu} }  \notag  \\
 	& 0 \le n_{j|2\nu} \le  N_{tot}\frac{2P_{\nu}^2 }{M} \tilde{P}_{j|2\nu} + \sqrt{3 \ln(1/\varepsilon_a)N_{tot}\frac{2P_{\nu}^2 }{M} \tilde{P}_{j|2\nu} }.   \notag \\
 	\label{linearprogramming}
 	\end{alignat}
 \end{small}
 
 Here, the last two constraints come from the simple fact that as the number of detection of $\tau_{j|2\mu}$, $n_{j|2\mu}$ cannot be larger than the number of preparing $\tau_{j|2\mu}$, which is around $N_{tot}2P_{\mu}^2\tilde{P}_{j|2\mu} /M $. For ease of calculation, in all these constraints  we let $\varepsilon_0 = \varepsilon_1 = \varepsilon_2 = \varepsilon_a$, then the total failture probability that we get the bound $\triangle^j_{\mu\nu}$ is $6\varepsilon_a$. Meanwhile, the failure probability that we get the last two bounds on is $2\varepsilon_a$. Hence, the total failture probability of all these constraints is $6(M+2)\varepsilon_a+2M \varepsilon_a= (8M+12) \varepsilon_a$.
 To conclude, with the help of the linear programming in \eqref{linearprogramming}, one can calculate the upper-bound of $n_{ph}$.

\bibliographystyle{plainnat}
\bibliography{references.bib}

\begin{thebibliography}{46}
\providecommand{\natexlab}[1]{#1}
\providecommand{\url}[1]{\texttt{#1}}
\expandafter\ifx\csname urlstyle\endcsname\relax
  \providecommand{\doi}[1]{doi: #1}\else
  \providecommand{\doi}{doi: \begingroup \urlstyle{rm}\Url}\fi

\bibitem[Abell{\'a}n et~al.(2014)Abell{\'a}n, Amaya, Jofre, Curty, Ac{\'\i}n,
  Capmany, Pruneri, and Mitchell]{abellan2014ultra}
C~Abell{\'a}n, W~Amaya, M~Jofre, M~Curty, A~Ac{\'\i}n, J~Capmany, V~Pruneri,
  and MW~Mitchell.
\newblock Ultra-fast quantum randomness generation by accelerated phase
  diffusion in a pulsed laser diode.
\newblock \emph{Optics express}, 22\penalty0 (2):\penalty0 1645--1654, 2014.

\bibitem[Ben-Or et~al.(2005)Ben-Or, Horodecki, Leung, Mayers, and
  Oppenheim]{ben2005universal}
Michael Ben-Or, Micha{\l} Horodecki, Debbie~W Leung, Dominic Mayers, and
  Jonathan Oppenheim.
\newblock The universal composable security of quantum key distribution.
\newblock In \emph{Theory of Cryptography Conference}, pages 386--406.
  Springer, 2005.

\bibitem[Bennett and Brassard(1984)]{bennett1984proceedings}
Charles~H Bennett and Gilles Brassard.
\newblock Proceedings of the ieee international conference on computers,
  systems and signal processing, 1984.

\bibitem[Boaron et~al.(2018)Boaron, Boso, Rusca, Vulliez, Autebert, Caloz,
  Perrenoud, Gras, Bussi{\`e}res, Li, et~al.]{boaron2018secure}
Alberto Boaron, Gianluca Boso, Davide Rusca, C{\'e}dric Vulliez, Claire
  Autebert, Misael Caloz, Matthieu Perrenoud, Ga{\"e}tan Gras, F{\'e}lix
  Bussi{\`e}res, Ming-Jun Li, et~al.
\newblock Secure quantum key distribution over 421 km of optical fiber.
\newblock \emph{Physical review letters}, 121\penalty0 (19):\penalty0 190502,
  2018.

\bibitem[Cao et~al.(2015)Cao, Zhang, Lo, and Ma]{cao2015discrete}
Zhu Cao, Zhen Zhang, Hoi-Kwong Lo, and Xiongfeng Ma.
\newblock Discrete-phase-randomized coherent state source and its application
  in quantum key distribution.
\newblock \emph{New Journal of Physics}, 17\penalty0 (5):\penalty0 053014,
  2015.

\bibitem[Chen et~al.(2020)Chen, Zhang, Liu, Jiang, Zhang, Hu, Guan, Yu, Xu,
  Lin, et~al.]{chen2020sending}
Jiu-Peng Chen, Chi Zhang, Yang Liu, Cong Jiang, Weijun Zhang, Xiao-Long Hu,
  Jian-Yu Guan, Zong-Wen Yu, Hai Xu, Jin Lin, et~al.
\newblock Sending-or-not-sending with independent lasers: Secure twin-field
  quantum key distribution over 509 km.
\newblock \emph{Physical review letters}, 124\penalty0 (7):\penalty0 070501,
  2020.

\bibitem[Chen et~al.(2021{\natexlab{a}})Chen, Zhang, Liu, Jiang, Zhang, Han,
  Ma, Hu, Li, Liu, et~al.]{chen2021twin}
Jiu-Peng Chen, Chi Zhang, Yang Liu, Cong Jiang, Wei-Jun Zhang, Zhi-Yong Han,
  Shi-Zhao Ma, Xiao-Long Hu, Yu-Huai Li, Hui Liu, et~al.
\newblock Twin-field quantum key distribution over a 511 km optical fibre
  linking two distant metropolitan areas.
\newblock \emph{Nature Photonics}, pages 1--6, 2021{\natexlab{a}}.

\bibitem[Chen et~al.(2021{\natexlab{b}})Chen, Zhang, Chen, Cai, Liao, Zhang,
  Chen, Yin, Ren, Chen, et~al.]{chen2021integrated}
Yu-Ao Chen, Qiang Zhang, Teng-Yun Chen, Wen-Qi Cai, Sheng-Kai Liao, Jun Zhang,
  Kai Chen, Juan Yin, Ji-Gang Ren, Zhu Chen, et~al.
\newblock An integrated space-to-ground quantum communication network over
  4,600 kilometres.
\newblock \emph{Nature}, 589\penalty0 (7841):\penalty0 214--219,
  2021{\natexlab{b}}.

\bibitem[Cui et~al.(2019)Cui, Yin, Wang, Chen, Wang, Guo, and Han]{cui2019twin}
Chaohan Cui, Zhen-Qiang Yin, Rong Wang, Wei Chen, Shuang Wang, Guang-Can Guo,
  and Zheng-Fu Han.
\newblock Twin-field quantum key distribution without phase postselection.
\newblock \emph{Physical Review Applied}, 11\penalty0 (3):\penalty0 034053,
  2019.

\bibitem[Curras~Lorenzo et~al.(2020)Curras~Lorenzo, Wooltorton, and
  Razavi]{curras2020twin}
Guillermo Curras~Lorenzo, Lewis Wooltorton, and Mohsen Razavi.
\newblock Twin-field quantum key distribution with fully discrete phase
  randomization.
\newblock \emph{Physical Review Applied}, 2020.

\bibitem[Curr{\'a}s-Lorenzo et~al.(2021)Curr{\'a}s-Lorenzo, Navarrete, Azuma,
  Kato, Curty, and Razavi]{curras2021tight}
Guillermo Curr{\'a}s-Lorenzo, {\'A}lvaro Navarrete, Koji Azuma, Go~Kato, Marcos
  Curty, and Mohsen Razavi.
\newblock Tight finite-key security for twin-field quantum key distribution.
\newblock \emph{npj Quantum Information}, 7\penalty0 (1):\penalty0 1--9, 2021.

\bibitem[Curty et~al.(2019)Curty, Azuma, and Lo]{curty2019simple}
Marcos Curty, Koji Azuma, and Hoi-Kwong Lo.
\newblock Simple security proof of twin-field type quantum key distribution
  protocol.
\newblock \emph{npj Quantum Information}, 5\penalty0 (1):\penalty0 1--6, 2019.

\bibitem[Dixon et~al.(2015)Dixon, Dynes, Lucamarini, Fr{\"o}hlich, Sharpe,
  Plews, Tam, Yuan, Tanizawa, Sato, et~al.]{dixon2015high}
AR~Dixon, JF~Dynes, M~Lucamarini, B~Fr{\"o}hlich, AW~Sharpe, A~Plews, S~Tam,
  ZL~Yuan, Y~Tanizawa, H~Sato, et~al.
\newblock High speed prototype quantum key distribution system and long term
  field trial.
\newblock \emph{Optics express}, 23\penalty0 (6):\penalty0 7583--7592, 2015.

\bibitem[Ekert(1991)]{ekert1991quantum}
Artur~K Ekert.
\newblock Quantum cryptography based on bell’s theorem.
\newblock \emph{Physical review letters}, 67\penalty0 (6):\penalty0 661, 1991.

\bibitem[Fang et~al.(2020)Fang, Zeng, Liu, Zou, Wu, Tang, Sheng, Xiang, Zhang,
  Li, et~al.]{fang2020implementation}
Xiao-Tian Fang, Pei Zeng, Hui Liu, Mi~Zou, Weijie Wu, Yan-Lin Tang, Ying-Jie
  Sheng, Yao Xiang, Weijun Zhang, Hao Li, et~al.
\newblock Implementation of quantum key distribution surpassing the linear
  rate-transmittance bound.
\newblock \emph{Nature Photonics}, pages 1--4, 2020.

\bibitem[Hwang(2003)]{hwang2003quantum}
Won-Young Hwang.
\newblock Quantum key distribution with high loss: toward global secure
  communication.
\newblock \emph{Physical Review Letters}, 91\penalty0 (5):\penalty0 057901,
  2003.

\bibitem[Ivanovic(1987)]{ivanovic1987differentiate}
Igor~D Ivanovic.
\newblock How to differentiate between non-orthogonal states.
\newblock \emph{Physics Letters A}, 123\penalty0 (6):\penalty0 257--259, 1987.

\bibitem[Koashi(2009)]{koashi2009simple}
M~Koashi.
\newblock Simple security proof of quantum key distribution based on
  complementarity.
\newblock \emph{New Journal of Physics}, 11\penalty0 (4):\penalty0 045018,
  2009.

\bibitem[Liao et~al.(2017)Liao, Cai, Liu, Zhang, Li, Ren, Yin, Shen, Cao, Li,
  et~al.]{liao2017satellite}
Sheng-Kai Liao, Wen-Qi Cai, Wei-Yue Liu, Liang Zhang, Yang Li, Ji-Gang Ren,
  Juan Yin, Qi~Shen, Yuan Cao, Zheng-Ping Li, et~al.
\newblock Satellite-to-ground quantum key distribution.
\newblock \emph{Nature}, 549\penalty0 (7670):\penalty0 43--47, 2017.

\bibitem[Lin and L{\"u}tkenhaus(2018)]{lin2018simple}
Jie Lin and Norbert L{\"u}tkenhaus.
\newblock Simple security analysis of phase-matching
  measurement-device-independent quantum key distribution.
\newblock \emph{Physical Review A}, 98\penalty0 (4):\penalty0 042332, 2018.

\bibitem[Liu et~al.(2019)Liu, Yu, Zhang, Guan, Chen, Zhang, Hu, Li, Jiang, Lin,
  et~al.]{liu2019experimental}
Yang Liu, Zong-Wen Yu, Weijun Zhang, Jian-Yu Guan, Jiu-Peng Chen, Chi Zhang,
  Xiao-Long Hu, Hao Li, Cong Jiang, Jin Lin, et~al.
\newblock Experimental twin-field quantum key distribution through sending or
  not sending.
\newblock \emph{Physical Review Letters}, 123\penalty0 (10):\penalty0 100505,
  2019.

\bibitem[Lo and Chau(1999)]{lo1999unconditional}
Hoi-Kwong Lo and Hoi~Fung Chau.
\newblock Unconditional security of quantum key distribution over arbitrarily
  long distances.
\newblock \emph{science}, 283\penalty0 (5410):\penalty0 2050--2056, 1999.

\bibitem[Lo et~al.(2005)Lo, Ma, and Chen]{lo2005decoy}
Hoi-Kwong Lo, Xiongfeng Ma, and Kai Chen.
\newblock Decoy state quantum key distribution.
\newblock \emph{Physical review letters}, 94\penalty0 (23):\penalty0 230504,
  2005.

\bibitem[Lo et~al.(2012)Lo, Curty, and Qi]{lo2012measurement}
Hoi-Kwong Lo, Marcos Curty, and Bing Qi.
\newblock Measurement-device-independent quantum key distribution.
\newblock \emph{Physical review letters}, 108\penalty0 (13):\penalty0 130503,
  2012.

\bibitem[Ma et~al.(2018)Ma, Zeng, and Zhou]{ma2018phase}
Xiongfeng Ma, Pei Zeng, and Hongyi Zhou.
\newblock Phase-matching quantum key distribution.
\newblock \emph{Physical Review X}, 8\penalty0 (3):\penalty0 031043, 2018.

\bibitem[Mayers(2001)]{10.1145/382780.382781}
Dominic Mayers.
\newblock Unconditional security in quantum cryptography.
\newblock \emph{J. ACM}, 48\penalty0 (3):\penalty0 351–406, May 2001.
\newblock ISSN 0004-5411.

\bibitem[Minder et~al.(2019)Minder, Pittaluga, Roberts, Lucamarini, Dynes,
  Yuan, and Shields]{minder2019experimental}
M~Minder, M~Pittaluga, GL~Roberts, M~Lucamarini, JF~Dynes, ZL~Yuan, and
  AJ~Shields.
\newblock Experimental quantum key distribution beyond the repeaterless secret
  key capacity.
\newblock \emph{Nature Photonics}, 13\penalty0 (5):\penalty0 334--338, 2019.

\bibitem[M{\"u}ller-Quade and Renner(2009)]{muller2009composability}
J{\"o}rn M{\"u}ller-Quade and Renato Renner.
\newblock Composability in quantum cryptography.
\newblock \emph{New Journal of Physics}, 11\penalty0 (8):\penalty0 085006,
  2009.

\bibitem[Pirandola et~al.(2017)Pirandola, Laurenza, Ottaviani, and
  Banchi]{pirandola2017fundamental}
Stefano Pirandola, Riccardo Laurenza, Carlo Ottaviani, and Leonardo Banchi.
\newblock Fundamental limits of repeaterless quantum communications.
\newblock \emph{Nature communications}, 8\penalty0 (1):\penalty0 1--15, 2017.

\bibitem[Pittaluga et~al.(2021)Pittaluga, Minder, Lucamarini, Sanzaro,
  Woodward, Li, Yuan, and Shields]{pittaluga2021600}
Mirko Pittaluga, Mariella Minder, Marco Lucamarini, Mirko Sanzaro, Robert~I
  Woodward, Ming-Jun Li, Zhiliang Yuan, and Andrew~J Shields.
\newblock 600-km repeater-like quantum communications with dual-band
  stabilization.
\newblock \emph{Nature Photonics}, pages 1--6, 2021.

\bibitem[Poppe et~al.(2008)Poppe, Peev, and Maurhart]{poppe2008outline}
Andreas Poppe, Momtchil Peev, and Oliver Maurhart.
\newblock Outline of the secoqc quantum-key-distribution network in vienna.
\newblock \emph{International Journal of Quantum Information}, 6\penalty0
  (02):\penalty0 209--218, 2008.

\bibitem[Sasaki et~al.(2011)Sasaki, Fujiwara, Ishizuka, Klaus, Wakui, Takeoka,
  Miki, Yamashita, Wang, Tanaka, et~al.]{sasaki2011field}
Masahide Sasaki, Mikio Fujiwara, H~Ishizuka, W~Klaus, K~Wakui, M~Takeoka,
  S~Miki, T~Yamashita, Z~Wang, A~Tanaka, et~al.
\newblock Field test of quantum key distribution in the tokyo qkd network.
\newblock \emph{Optics express}, 19\penalty0 (11):\penalty0 10387--10409, 2011.

\bibitem[Shor and Preskill(2000)]{shor2000simple}
Peter~W Shor and John Preskill.
\newblock Simple proof of security of the bb84 quantum key distribution
  protocol.
\newblock \emph{Physical review letters}, 85\penalty0 (2):\penalty0 441, 2000.

\bibitem[Takeoka et~al.(2014)Takeoka, Guha, and Wilde]{takeoka2014fundamental}
Masahiro Takeoka, Saikat Guha, and Mark~M Wilde.
\newblock Fundamental rate-loss tradeoff for optical quantum key distribution.
\newblock \emph{Nature communications}, 5\penalty0 (1):\penalty0 1--7, 2014.

\bibitem[Tamaki et~al.(2018)Tamaki, Lo, Wang, and
  Lucamarini]{tamaki2018information}
Kiyoshi Tamaki, Hoi-Kwong Lo, Wenyuan Wang, and Marco Lucamarini.
\newblock Information theoretic security of quantum key distribution overcoming
  the repeaterless secret key capacity bound.
\newblock \emph{arXiv preprint arXiv:1805.05511}, 2018.

\bibitem[Wang et~al.(2020)Wang, Yin, Lu, Wang, Chen, Zhang, Huang, Xu, Guo, and
  Han]{wang2020optimized}
Rong Wang, Zhen-Qiang Yin, Feng-Yu Lu, Shuang Wang, Wei Chen, Chun-Mei Zhang,
  Wei Huang, Bing-Jie Xu, Guang-Can Guo, and Zheng-Fu Han.
\newblock Optimized protocol for twin-field quantum key distribution.
\newblock \emph{Communications Physics}, 3\penalty0 (1):\penalty0 1--7, 2020.

\bibitem[Wang et~al.(2014)Wang, Chen, Yin, Li, He, Li, Zhou, Song, Li, Wang,
  et~al.]{wang2014field}
Shuang Wang, Wei Chen, Zhen-Qiang Yin, Hong-Wei Li, De-Yong He, Yu-Hu Li, Zheng
  Zhou, Xiao-Tian Song, Fang-Yi Li, Dong Wang, et~al.
\newblock Field and long-term demonstration of a wide area quantum key
  distribution network.
\newblock \emph{Optics express}, 22\penalty0 (18):\penalty0 21739--21756, 2014.

\bibitem[Wang et~al.(2019)Wang, He, Yin, Lu, Cui, Chen, Zhou, Guo, and
  Han]{wang2019beating}
Shuang Wang, De-Yong He, Zhen-Qiang Yin, Feng-Yu Lu, Chao-Han Cui, Wei Chen,
  Zheng Zhou, Guang-Can Guo, and Zheng-Fu Han.
\newblock Beating the fundamental rate-distance limit in a proof-of-principle
  quantum key distribution system.
\newblock \emph{Physical Review X}, 9\penalty0 (2):\penalty0 021046, 2019.

\bibitem[Wang(2005)]{wang2005beating}
Xiang-Bin Wang.
\newblock Beating the photon-number-splitting attack in practical quantum
  cryptography.
\newblock \emph{Physical review letters}, 94\penalty0 (23):\penalty0 230503,
  2005.

\bibitem[Wang et~al.(2018)Wang, Yu, and Hu]{wang2018twin}
Xiang-Bin Wang, Zong-Wen Yu, and Xiao-Long Hu.
\newblock Twin-field quantum key distribution with large misalignment error.
\newblock \emph{Physical Review A}, 98\penalty0 (6):\penalty0 062323, 2018.

\bibitem[Xu et~al.(2012)Xu, Qi, Ma, Xu, Zheng, and Lo]{xu2012ultrafast}
Feihu Xu, Bing Qi, Xiongfeng Ma, He~Xu, Haoxuan Zheng, and Hoi-Kwong Lo.
\newblock Ultrafast quantum random number generation based on quantum phase
  fluctuations.
\newblock \emph{Optics express}, 20\penalty0 (11):\penalty0 12366--12377, 2012.

\bibitem[Yin and Fu(2019)]{yin2019measurement}
Hua-Lei Yin and Yao Fu.
\newblock Measurement-device-independent twin-field quantum key distribution.
\newblock \emph{Scientific reports}, 9\penalty0 (1):\penalty0 1--13, 2019.

\bibitem[Yin et~al.(2016)Yin, Chen, Yu, Liu, You, Zhou, Chen, Mao, Huang,
  Zhang, et~al.]{yin2016measurement}
Hua-Lei Yin, Teng-Yun Chen, Zong-Wen Yu, Hui Liu, Li-Xing You, Yi-Heng Zhou,
  Si-Jing Chen, Yingqiu Mao, Ming-Qi Huang, Wei-Jun Zhang, et~al.
\newblock Measurement-device-independent quantum key distribution over a 404 km
  optical fiber.
\newblock \emph{Physical review letters}, 117\penalty0 (19):\penalty0 190501,
  2016.

\bibitem[Zhang et~al.(2020)Zhang, Xu, Wang, and Wang]{PhysRevApplied.14.064070}
Chun-Mei Zhang, Yi-Wei Xu, Rong Wang, and Qin Wang.
\newblock Twin-field quantum key distribution with discrete-phase-randomized
  sources.
\newblock \emph{Phys. Rev. Applied}, 14:\penalty0 064070, Dec 2020.

\bibitem[Zhang et~al.(2017)Zhang, Zhao, Razavi, and Ma]{zhang2017improved}
Zhen Zhang, Qi~Zhao, Mohsen Razavi, and Xiongfeng Ma.
\newblock Improved key-rate bounds for practical decoy-state
  quantum-key-distribution systems.
\newblock \emph{Physical Review A}, 95\penalty0 (1):\penalty0 012333, 2017.

\bibitem[Zhong et~al.(2019)Zhong, Hu, Curty, Qian, and Lo]{zhong2019proof}
Xiaoqing Zhong, Jianyong Hu, Marcos Curty, Li~Qian, and Hoi-Kwong Lo.
\newblock Proof-of-principle experimental demonstration of twin-field type
  quantum key distribution.
\newblock \emph{Physical Review Letters}, 123\penalty0 (10):\penalty0 100506,
  2019.

\end{thebibliography}
\end{document}